\def\thin{{\thinspace}}

\def\ref{\par \noindent \hangindent=3pc \hangafter=1}

\def\sep{{\par \noindent \hangindent=15pt \hangafter=1}}

\def\scr{\scriptstyle}
\def\etal{{{\it et al}.\ }}

\def\pn{{\par\noindent}}
\def\ie{{{\it i.e}.}}
\def\eq{{\ \equiv\ }}

\def\rbar{{\overline{r}}}
\def\rhobar{\lower 0.2ex\hbox{${\overline{\rho}}$}}
\def\equ{\  = \ }

\def\;{;\thin}
\def\vecomega{{\rlap{$\omega$}{\hskip 0.1ex\hbox{$\omega$}}}}

\def\vecOmega{{\rlap{$\Omega$}{\hskip 0.1ex\hbox{$\Omega$}}}}

\def\vectheta{{\rlap{$\theta$}{\hskip 0.1ex\hbox{$\theta$}}}}
\def\thetahat{{\rlap{\vectheta}{\hskip 0.2ex{\raise 0.5ex\hbox{\overline}}}}     }

\def\Omegadot{{\hbox{$\dot \Omega$}}}

\def\Pdot{{\hbox{$\dot P$}}}

\def\edot{{\hbox{$\dot e$}}}

\def\adot{{\hbox{$\dot a$}}}

\def\omegadot{{\hbox{$\dot \omega$}}}

\def\dbfdot{{\hbox{$\dot {\bf d}$}}}

\def\dbfddot{{\hbox{$\ddot {\bf d}$}}}

\def\h{{{\bf h}}}
\def\e{{{\bf e}}}
\def\q{{{\bf q}}}
\def\f{{{\bf f}}}
\def\bd{{{\bf d}}}

\def\r{{{\bf r}}}
\def\B{{{\bf B}}}

\def\hhat{{\overline{\bf h}}}
\def\ehat{{{\overline{\bf e}}}}
\def\qhat{{{\overline{\bf q}}}}
\def\Hhat{{{\overline{\bf H}}}}

\def\w{{\ \ \ \ \ }}

\def\tgs{{\thin \rlap{\raise 0.5ex\hbox{$\scr  {>}$}}{\lower 0.3ex\hbox{$\scr  {\sim}$}} \thin }}
\def\tls{{\thin \rlap{\raise 0.5ex\hbox{$\scr  {<}$}}{\lower 0.3ex\hbox{$\scr  {\sim}$}} \thin }}
\def\tll{{\raise 0.3ex\hbox{$\scr  {\thin \ll \thin }$}}}
\def\tgg{{\raise 0.3ex\hbox{$\scr  {\thin \gg \thin }$}}}
\def\tle{{\raise 0.3ex\hbox{$\scr  {\thin \le \thin }$}}}
\def\tge{{\raise 0.3ex\hbox{$\scr  {\thin \ge \thin }$}}}
\def\tl{{\raise 0.3ex\hbox{$\scr  {\thin < \thin }$}}}
\def\tg{{\raise 0.3ex\hbox{$\scr  {\thin > \thin }$}}}
\def\ts{{\raise 0.3ex\hbox{$\scr  {\thin \sim \thin }$}}}
\def\ets{{\ \ts\ }}

\def\tp{{\raise 0.3ex\hbox{\fiverm +}}}
\def\Chi{{\raise 0.4ex\hbox{$\chi$}}}

\def\deg{{^\circ}}

\def\K{{\rm\thin K}}

\def\Msun{\hbox{$\thin M_{\odot}$}}

\def\Lsun{\hbox{$\thin L_{\odot}$}}

\def\Rsun{\hbox{$\thin R_{\odot}$}}

\def\p{\thin\tp\thin}

\input psfig.sty
\documentclass[preprint]{aastex}
\newcommand{\be}{\begin{equation}}
\newcommand{\ee}{\end{equation}}
\begin{document}
\shorttitle{Orbital Evolution}
\title{Orbital Evolution in Binary and Triple Stars, with an application to SS Lac}
\author{Peter P. Eggleton\altaffilmark{1,2} \& Ludmila Kiseleva-Eggleton\altaffilmark{1}}
\altaffiltext{1}{Lawrence Livermore National Laboratory, Livermore, CA 94550\\
    Email: {\tt ppe@igpp.ucllnl.org, lkisseleva@igpp.ucllnl.org}}
\altaffiltext{2}{On leave from the Institute of Astronomy, Madingley Rd,
         Cambridge CB3 0HA, UK}

\def\rbar{{{r_*}}}
\def\k{{\bf k}}
\def\kdot{{\hbox{$\dot k$}}}
\def\kbfdot{{\hbox{$\dot {\bf k}$}}}
\def\r{{{\bf r}}}
\def\EEdot{\hbox{$\dot E$}}
\def\vbf{{{\bf v}}}
\def\vO{\vecOmega}
\def\et{equilibrium-tide\ }
\def\tf{tidal friction\ }
\def\smc{0045-7319}
\def\Te{T_{\rm E}}
\def\spc{\hskip 0.3truein}
\def\p{^{\prime}}
\def\olp{\omega_{\rm lp}}
\def\olpA{\omega_{\rm lp,A}}
\def\K{{\bf K}}
\def\H{{\bf H}}
\def\E{{\bf E}}
\def\Q{{\bf Q}}
\def\Jhat{\overline{\bf J}} 
\def\uhat{{\overline{\bf u}}}
\def\Hbar{{\overline{H}}}
\def\alpdot{{d\aJ\over dt}}
\def\olpdot{{\hbox{$\dot \omega_{\rm lp}$}}}
\def\tTF{t_{\rm F}}
\def\tTFa{t_{\rm F1}}
\def\Oe{\Omega_{\rm e}}
\def\Oq{\Omega_{\rm q}}
\def\Oh{\Omega_{\rm h}}
\def\Oae{\Omega_{\rm 1e}}
\def\Oaq{\Omega_{\rm 1q}}
\def\Oah{\Omega_{\rm 1h}}
\def\Obe{\Omega_{\rm 2e}}
\def\Obq{\Omega_{\rm 2q}}
\def\Obh{\Omega_{\rm 2h}}
\def\He{\overline{H}_{\rm e}}
\def\Hq{\overline{H}_{\rm q}}
\def\Hh{\overline{H}_{\rm h}}
\def\ZGR{{Z_{\rm GR}}}
\def\tv{{t_{\rm V}}}
\def\tva{{t_{\rm V1}}}
\def\Aa{_{\rm Aa}}
\def\Ab{_{\rm Ab}}
\def\B{_{\rm B}}
\def\AB{_{\rm AB}}
\def\A{_{\rm A}}
\def\dfII{\Delta\Phi_{\rm II}}
\def\eout{{e_{\rm out}}}
\def\aH{{\alpha_{\rm H}}}
\def\bH{{\beta_{\rm H}}}
\def\aJ{{\alpha_{\rm J}}}
\def\bJ{{\beta_{\rm J}}}
\def\aOa{{\alpha_{\Omega_1}}}
\def\bOa{{\beta_{\Omega_1}}}
\def\aOb{{\alpha_{\Omega_2}}}
\def\bOb{{\beta_{\Omega_2}}}
\def\Prot{{P_{\rm rot}}}
\def\alpdot{\dot{\aJ}}
\def\betdot{\dot{\bJ}}
\def\vr{{V_{\rm rot}}}

\begin{abstract} 
Abstract:  We present equations governing the way in which both the orbit
and the intrinsic spins of stars in a close binary should evolve subject
to a number of perturbing forces, including the effect of a third body in
a possibly inclined wider orbit. We illustrate the solutions in some
binary-star and triple-star situations: tidal friction in a wide but eccentric
orbit of a radio pulsar about a B star (\smc), the Darwin and eccentricity
instabilities in a more massive but shorter-period massive X-ray binary,
and the interaction of tidal friction with Kozai cycles in a triple such
as $\beta$ Per, at an early stage in that star's life when all 3 components
were ZAMS stars. We also attempt to model in some detail
the interesting triple system SS Lac, which stopped eclipsing in about
1950. We find that our model of SS Lac is quite constrained by the relatively
good observational data of this system, and leads to a specific
inclination $(29\deg$) of the outer orbit relative to the inner orbit at epoch
zero (1912). Although the intrinsic spins of the stars have little
effect on the orbit, the converse is not true: the spin axes can vary
their orientation relative to the close binary by up to $120\deg$ on a 
timescale of about a century.
\end{abstract}

\keywords{binary stars; triple stars; tidal friction}


\section{Introduction}

We model the effect on a short-period binary-star orbit, and also on the spins of
the two components, of the following perturbations:
\sep(a) a third body (treated as a point mass) in a longer-period orbit
\sep(b) the quadrupolar distortion of the stars due to their intrinsic spin
\sep(c) the further quadrupolar distortion due to their mutual gravity
\sep(d) tidal friction, in the equilibrium-tide approximation
\sep(e) General Relativity.
\pn The third body's effect is treated only at the quadrupole level of 
approximation, although in principle it should not be difficult to go to
a higher order if necessary. The stellar distortion terms are also treated 
only at the quadrupole level; it might be rather harder here to go to a
higher order, but it is probably even less necessary. Each of the
perturbing forces has been averaged over an approximately Keplerian orbit.
We follow the analysis of Eggleton, Kiseleva \& Hut (1998) -- hereinafter 
EKH98 -- for effects (b) -- (d). The third-body perturbation comes from the 
same type of analysis (as does the familiar Schwarzschild-metric correction).

We illustrate the model with some binary and triple examples: 
\sep (i) the 
circularisation of an initially eccentric orbit of a neutron star around an
obliquely-rotating massive normal B star, based on the SMC radio-pulsar \smc
\ (Kaspi \etal 1994);
we assume that the B star has spin inclined at a large angle ($135\deg$) to
the orbit, and model the way in which the spin parallelises as well as 
pseudo-synchronises with the orbit
\sep (ii) the Darwin instability, i.e. the tendency for an 
orbit to {\it de-}synchronise if the spin angular momentum of a star is more 
than a third of the orbital angular momentum; and the eccentricity instability,
in which rapid enough prograde rotation of the star causes the eccentricity
to increase
\sep (iii) Kozai cycles, i.e. 
cyclic large-amplitude variation of the eccentricity of the inner binary
binary due to a highly-inclined outer orbit, which in combination 
with tidal friction can make the inner orbit shrink
even if it is initially too wide for tidal friction to be important. 
\pn We base some of these illustrations on actual stellar systems, but the 
data on these systems is not (yet) sufficient to test the model rigorously.

We then apply the model to the 
interesting $14.4\thin$d binary SS~Lac, which eclipsed for the first 50 years 
of the 20th century, and stopped eclipsing later. Recently Torres \& Stefanik 
(2000) -- hereinafter TS00 -- have demonstrated the existence of
a third body in a $\ts 700\thin$d orbit, which if it is inclined at a suitable
angle to the inner orbit could well account for the necessary change in
the orientation of the orbit relative to the observer. We find that in order
to accommodate the data given by TS00, we require the outer orbit to be inclined 
at about $29\deg$ (or 151$\deg$) to the inner orbit. We also require
a specific value ($37\deg$) for the longitude of the outer orbit's axis
relative to the inner orbital frame in 1912 (epoch zero).
We find a modest variation with time in the eccentricity, and in the rate 
of change of inclination of the inner orbit to the 
line of sight. These cause us to revise slightly the masses found by TS00.
Our model is fully constrained by the known data on this system, but it is not
over-constrained, so that unfortunately we are not in a position to confirm 
that the model is correct. We can however make predictions for changes that
should be capable of confirmation or refutation in about 20 years.

In \S2, we set out the equations governing the change in the orbit (and
also in the rates of rotation of the two components), discussing the
level of approximation that we use. In \S3 we illustrate the behaviour
of the equations in a number of straightforward cases, two without and
one with a third body. In \S4 we apply our model to SS Lac, and we conclude 
with a discussion in \S5. Some mathematical details are given in an Appendix.

\section{Equations for orbital evolution}

The evolution of the orbit under the influence of the perturbations listed in \S1
is well expressed in terms of the following 5 vectors: $\e$, the Laplace-Runge-Lenz
vector, which points along the major axis in the direction of periastron
and has magnitude $e$, the eccentricity; $\h$, the orbital angular momentum
vector, pointing perpendicular to the orbital plane and with magnitude
$h$, the orbital angular momentum per unit reduced mass $\mu$; $\q\eq\h\wedge\e$,
which makes a right-handed orthogonal triad $\e,\thin\q,\thin\h$ with the
previous two vectors, since $\e$ and $\h$ are always mutually perpendicular; 
and also the spin vectors $\vO_1,\thin\vO_2$ of the two components. The vector 
$\q$ is along the latus rectum, the line through the focus
parallel to the minor axis. It is also convenient to define unit vectors 
$\ehat,\thin\qhat,\thin\hhat$, a right-handed orthogonal unit basis.
This basis is not an inertial frame, of course, since it varies with time
as the system evolves under the perturbations. But provided that the 
perturbations are sufficiently small that they do not affect the orbit
by more than a small amount on timescales less than the period of the
outer orbit, it is possible to estimate rather simply the rates of change of 
these 5 vectors in response to the 5 perturbative forces listed above.

The equations (EKH98) governing the rates of change of
$\e,\thin\h,\thin\vO_1,\thin\vO_2$ can be written as follows:

$${1\over e}\thin{d\e\over dt}\equ (Z_1+Z_2+\ZGR)\thin\qhat-(Y_1+Y_2)\thin\hhat
-(V_1+V_2)\ehat\hskip 1truein$$
$$\hskip 0.7truein-(1-e^2)\thin\{5S_{eq}\ehat-(4S_{ee}-S_{qq})\thin\qhat+S_{qh}\thin\hhat\}\w,
\eqno(1)$$

$${1\over h}\thin{d\h\over dt}\equ (Y_1+Y_2)\thin\ehat-(X_1+X_2)\thin\qhat-
(W_1+W_2)\thin\hhat\hskip 1truein$$
$$\hskip 0.7truein  +(1-e^2)S_{qh}\thin\ehat-(4e^2+1)S_{eh}\thin\qhat+5e^2S_{eq}
\thin\hhat\w,\eqno(2)$$

$$I_1{d\vO_1\over dt}\equ \mu h\{-Y_1\thin\ehat+X_1\thin\qhat+W_1\thin\hhat\}\w,
\hskip 1truein\eqno(3)$$

$$I_2{d\vO_2\over dt}\equ \mu h\{-Y_2\thin\ehat+X_2\thin\qhat+W_2\thin\hhat\}\w.
\hskip 1truein\eqno(4)$$
An equation for $\q$ follows from differentiating the product $\q=\h\wedge\e$.
The quantities 
$\thin V_i,\thin W_i,\thin X_i,\thin Y_i,\thin Z_i$, one for each component of the inner 
pair, are given below. They arise from the quadrupolar distortion of the two components.
The first two are dissipative terms, due to tidal friction, which tend to enforce
orbital circularisation and synchronous rotation, at least in the absence of a third-body 
perturbation. The next three are mainly non-dissipative perturbations due to quadrupole
distortions, giving precession and apsidal motion; but $X,\thin Y$ do contain
a small dissipative contribution which tends to bring the stellar rotations into
parallel with the orbit. The $Z$ term, giving apsidal motion, contains a GR correction.
The tensor $S$, with components $S_{ee},\thin S_{eq},\dots$ in the $\e,\thin\q,\thin\h$
frame, is due only to the third body, and its components are also given somewhat 
further below -- equations (15), (16). 

The dissipative terms $V_1,\thin W_1$ are
$$V_1\equ {9\over \tTFa}\left[ {1+{15\over 4}e^2+{15\over 8}e^4+{5\over 64}e^6 
\over (1-e^2)^{13/2}}\ -\ {11\Oah\over 18\omega}{1+{3\over 2}e^2+{1\over 8}e^4 
\over (1-e^2)^5}\right] \eqno(5)$$
$$W_1\equ {1\over \tTFa}\left[{1+{15\over 2}e^2+{45\over 8}e^4+{5\over 16}
e^6\over (1-e^2)^{13/2}}-{\Oah\over\omega}{1+3e^2+{3\over 8}e^4\over 
(1-e^2)^5}\right]\eqno(6)$$ 
with similar expressions for $*2$. $\Oah\eq\vO_1.\hhat$ is the component of $\vO_1$ in the direction
of $\hhat$, i.e. parallel to the orbital axis, and $\omega$ is the mean angular
velocity of the inner orbit, i.e. $2\pi/P$. The tidal-friction timescale 
$\tTF$ is estimated here, in terms of an inherent viscous timescale $\tv$ for
each star, as
$${1\over\tTFa}\equ {9\over\tva}
\thin{R_1^8\over a^8}\thin{MM_2\over M_1^2}\thin{1\over(1-Q_1)^2}\w. \eqno(7)$$
$M_1,\thin M_2$ are the masses of the two components of the inner binary, $M$ their 
sum, $R_1,\thin R_2$ their radii, and $a$ the semimajor axis. $Q_1$ is a coefficient
measuring the quadrupolar deformability of the star; it is closely related to the
apsidal motion constant (EKH98). For an $n\ts 3$ polytrope, 
$$I\equ 0.08MR^2\w,\w Q\equ 0.028\w.\eqno(8)$$

The intrinsic viscous timescale of the star, $\tva$, 
is not easily determined, but we use an estimate based on (a) the timescale on which the
star would be turned over if most of the luminosity $L_1$ were carried by convection
(Zahn Eggleton, 1977), and
(b) a dimensionless factor $\gamma$ which comes from integrating over the star the 
rate-of-strain tensor (squared) of the time-dependent tidal velocity field:
$${1\over\tva}\equ \gamma_1\left({L_1\over 3M_1R_1^2}\right)^{1/3}\w,\w\gamma_1\ts 0.01\w.
\eqno(9)$$
The timescale $\tva$ is of the order of years or decades. The quantity $\gamma$ -- see Appendix -- is 
determined by a model of the tidal amplitude as a function of radius through the star
(EKH98), obtained by solving explicitly the velocity field required by the continuity
equation if isobaric surfaces within the star are always to be equipotential
surfaces -- the basic assumption of the \et\ model.

The contributions $X_1,\thin Y_1,\thin Z_1$ to the rotation of the axes due to rotational 
and tidal 
distortion of $*1$ (including the small contribution of tidal friction), are given by
$$X_1\equ -{M_2A_1\over 2\mu\omega a^5}\thin{\Oah\thin\Oae\over(1-e^2)^2}-{\Oaq\over 2\omega 
\tTFa}\thin{1+{9\over 2}e^2+{5\over 8}e^4\over (1-e^2)^5}\w,\eqno(10)$$
$$Y_1\equ -{M_2A_1\over 2\mu\omega a^5}\thin{\Oah\thin\Oaq\over(1-e^2)^2}+{\Oae\over 2\omega 
\tTFa}\thin{1+{3\over 2}e^2+{1\over 8}e^4\over (1-e^2)^5}\w,\eqno(11)$$

$$Z_1\equ {M_2A_1\over2\mu\omega a^5}\left[{2\Oah^2-\Oae^2-
\Oaq^2\over2(1-e^2)^2}+{15GM_2 \over a^3}\thin{1+{3\over 2}e^2+
{1\over 8}e^4\over (1-e^2)^5}\right].\eqno(12)$$
Here $\mu$ is the reduced mass, and $\Oae,\Oaq$ are the components
of $\vO_1$ in the directions of $\ehat,\thin\qhat$. EKH98 give the
linear combinations $X_1\Oaq -Y_1\Oae$ and $X_1\Oae+Y_1\Oaq$, rather than $X_1,Y_1$ directly:
see the Appendix to this paper. The coefficient $A_1$ is
$$A_1\equ {R_1^5Q_1\over 1-Q_1}\w.\eqno(13)$$
In all of equations (5) -- (13), we interchange suffices 1 and 2 to find the corresponding
term for the second component of the inner binary.

The GR contribution to apsidal motion is
$$Z_{\rm GR}\equ {3 GM\omega\over ac^2(1-e^2)} \ \  .\eqno(14)$$

The effect of a third body (mass $M_3$) is included here only at the quadrupole level of 
approximation, following Kiseleva, Eggleton \& Mikkola 1998 (hereinafter KEM98).
At this level, the CG of the inner binary, and the third body, are unperturbed, and so the
outer orbit is exactly constant. Like the inner orbit it can be described by a right-handed
triad {\bf E,\thin Q,\thin H}. Within the inner binary, there is a perturbative force which in
the lowest approximation is linear in the vector separation: $\delta f_i\propto T_{ij}d_j$,
where $\bd$ is the separation of the inner pair. The tensor $T$ depends on {\bf D}, the 
separation of the outer pair (i.e. of the third body and the CG of the inner pair). Averaging 
$T$ over an outer orbit, and then averaging the effect of $\delta {\bf f}$ over the inner
orbit, assuming that {\it both} orbits are only slowly varying on this 
timescale, we find that the tensor $S$ of equations (1) and (2) is
$$S_{ij}\equ C\thin(\delta_{ij}-3\Hbar_i\Hbar_j)\w,\hskip1.2truein$$
$$\w\w C\equ  {M_3\thin\omega_{\rm out}^2
\over 4(M+M_3)\thin\omega\thin(1-e^2)^{1/2}\thin(1-e_{\rm out}^2)^{3/2}}\w.\eqno(15)$$
$\vert{\bf E}\vert\eq e_{\rm out}$ is the eccentricity of the outer orbit, and $\omega_{\rm out}$
is the outer orbital frequency.
Then the effect of the force $\delta\f$ on the vectors $\e,\thin\h$ are as indicated in
equations (1) -- (2), where on referring to the basis vectors $\e,\thin\q,\thin\h$ we
have
$$S_{ee}\equ C(1-3\thin\He\thin\He)\w,\w S_{eq}\equ -3C\thin\He\thin\Hq\ ,\ etc.\eqno(16)$$
A somewhat surprising but welcome simplification is that after averaging $T$ over {\bf D} 
we have dependence only on $\H$, and not on $\E,\thin \Q$ as well. Note that $C$ is {\it not}
a constant, because of its dependence on $e$.

Equations (1) -- (4) are closed by the fact that $a,\thin\omega$, which
appear in several places on the RHSs, are 
obtained in terms of $e,\thin h$ by way of the relations
$$a={h^2\over GM(1-e^2)}\w,\w \omega^2\equ\left({2\pi\over P}\right)^2\equ 
{GM\over a^3}\w.\eqno(17)$$
As $\ehat$ and $\hhat$ evolve away from their initial
values, along with $\vO_1,\vO_2$, we have to compute various vector and
scalar products: $\q\equ\h\wedge\e$, and $\Oe,\thin \Oq,\thin\He$, etc. 

The physical significance of $V,\thin W,\thin X,\thin Y,\thin Z$ is perhaps most 
easily seen by splitting each
of equations (1) and (2) into two pieces, one each for the moduli ($e,h$) and one each
for the unit vectors $\ehat,\thin \hhat$:
$${1\over e}\thin{de\over dt}\equ -V_1-V_2-5(1-e^2)S_{eq}\w,\hskip 0.8truein\eqno(18)$$
$${d\ehat\over dt}\equ \{Z_1+Z_2+\ZGR+(1-e^2)\thin(4S_{ee}-S_{qq})\}\thin\qhat\hskip 0.5truein$$
$$-\{Y_1+Y_2+(1-e^2)S_{qh}\}\thin\hhat\w,\eqno(19)$$
$${1\over h}\thin{dh\over dt}\equ -W_1-W_2+5e^2S_{eq}\w,\hskip 1truein\eqno(20)$$
$${d\hhat\over dt}\equ\{Y_1+Y_2+(1-e^2)S_{qh}\}\thin\ehat\hskip 1.2truein$$
$$-\{X_1+X_2+(4e^2+1)S_{eh}\}\thin\qhat\w.\eqno(21)$$
Equations (19) and (21) can both be written as
$${d\uhat\over dt}\equ\K\wedge\uhat\w,\eqno(22)$$
$$\K\equ(X_1+X_2+X_{\rm TB},\ Y_1+Y_2+Y_{\rm TB},\ Z_1+Z_2 +\ZGR+Z_{\rm TB})\ .\eqno(23)$$
Clearly $\K\eq(X,Y,Z)\thin\eq \thin X\ehat+Y\qhat+Z\hhat$ is the 
angular velocity of the $\e,\q,\h$ frame relative to an inertial frame. The terms
with suffix TB arise from the third body, and can be readily identified with the
corresponding $S$-terms in equations (19) and (21).
It is easy to see that $\qhat$ satisfies the same equation (22) as $\ehat,\thin\hhat$.

Equations (1) -- (4) can be integrated numerically, using for example a four-stage Runge-Kutta
procedure. However equation (1) as it stands has the slight problem, numerically, 
that in situations where $e\to 0$ (usually as a result of tidal friction), $\ehat$ 
becomes undefined. This is easily solved by using instead equations (18) and (19).
There is of course some redundancy in the $\ehat$ equation, but equations (18), (19)
together are very well-behaved. There is not the same problem with equation (2), 
since $h$ can hardly get to zero in realistic circumstances. Consequently, equations 
(2) -- (4), with (18) and (19), i.e. 13 first-order ODEs in all, are quite readily
integrated numerically as they stand.  There are in fact two redundancies, since
$\ehat.\thin\h=0$ as well as $\ehat.\thin\ehat = 1$.

We find it convenient to use as our computational (and inertial) frame 
the {\it initial} ($t=0$) orbital frame, say $\ehat_0,\thin\qhat_0,\thin\hhat_0$. 
We need to be given a number of constant scalars and vectors, i.e. $M_i,
\thin R_i,\thin L_i,\thin I_i,\thin Q_i$ for each of the inner pair of stars, 
and $M_3,\thin \H,\thin e_{\rm out}$ for the third body and outer orbit.
We also have to supply 13 initial values, for $e,\thin\ehat,\thin\h,
\thin\vO_1,\thin \vO_2$. Some of the components are rather trivial, because by the 
above definition $\ehat_0=(1,0,0)$ and $\hhat_0=(0,0,1)$ at $t=0$ -- and 
of course $\qhat_0=(0,1,0)$. The quantities $a,\thin\omega$ at $t=0$ follow
from $e,\thin h$ at $t=0$ by equations (17).

The non-trivial initial vectors $\H,\thin\vO_1,\thin \vO_2$ are all given 
directions in the obvious spherical-polar form, e.g. 
$$\Hhat\equ\sin\aH\cos\bH\thin\ehat+\sin\aH\sin\bH\thin\qhat+
\cos\aH\thin\hhat\w.\eqno(24)$$ 
Thus $\aH$ and $\bH$ are two of the three polar coordinates, the colatitude 
and longitude, of the vector $\Hhat$ in the orbital frame. We view $\H$ as 
a vector starting at the focus, and intersecting a sphere which is centred 
on the focus and has North pole on the $\hhat$ axis. Longitude zero (on the
equator) is on the $\ehat$ axis, \ie\ the projection of periastron. The 
components of $\Hhat$ are constant in the computational
frame of $\ehat_0,\thin\qhat_0,\thin\hhat_0$, but the components $\He,\thin\Hq,
\thin\Hh$ which appear in equations (16) change with time because 
$\e,\thin\q,\thin\h$ change with time in the $\ehat_0,
\thin\qhat_0,\thin\hhat_0$ frame, according to equation (22). 

The directions of $\vO_1,\thin\vO_2$ are given similarly, by pairs of angles 
$\aOa,\thin\bOa$, etc. These angles have to be given initially, but 
of course the $\vecOmega$'s, unlike $\H$, vary both in the inertial 
frame and in the instantaneous orbital frame.

In order to be able to determine the radial velocity curve and/or the
eclipse light curve, and how they might change with time, we have to 
specify in addition the (constant) direction, $\Jhat$, from which the 
orbit is observed -- constant (like $\Hhat$) in the 
$\ehat_0,\thin\qhat_0,\thin\hhat_0$ frame, but not of course in the 
$\ehat,\thin\qhat,\thin\hhat$ frame. We specify $\Jhat$ by two angles 
in the same way as $\overline{\bf H}$, and call them $\aJ,\thin\bJ$. The 
angle $\aJ$ is just the usual inclination of the orbit to the line of 
sight. The angle $\bJ$ is almost the same as the `longitude of periastron'. 
The latter quantity is usually called $\omega$, but we call it $\olp$ as we 
have already used $\omega$ for the orbital frequency. The relation between $\bJ$ and 
$\olp$ is
$$\bJ+\olp\equ 270\deg\w.\eqno(25)$$
We think of $\aJ$, $\bJ$ firstly as given initial conditions; but at later 
times they can be evaluated from
$$\cos\aJ\equ\Jhat.\thin\hhat\w,\w\tan\bJ\equ {\Jhat.\thin\qhat\over
\Jhat.\thin\ehat}\w.\eqno(26)$$
$\Jhat$ is a constant in space, but variable in the orbital frame
$\ehat,\thin\qhat,\thin\hhat$ since these unit vectors vary with time. 
The same formulae (26), {\it mutatis mutandis}, give the corresponding $\alpha$
(colatitude) and $\beta$ (longitude) for the other vectors $\H$, $\vO_1$, 
and $\vO_2$ at later times.

If we are fortunate enough to have very accurate observations over sufficiently
long stretches of time, we may be able to measure some rates of change such
as $\Pdot,\thin\edot,\thin \alpdot,\thin \betdot$. Equation (18) already gives
$\edot$, and along with equations (17) and (20) this gives $\Pdot$ (or 
$\omegadot$ or $\adot$):
$$-{\adot\over 2a}\equ{\omegadot\over 3\omega}\equ-{\Pdot\over 3P}
\equ W_1+W_2+{(V_1+V_2)e^2\over 1-e^2}\w.\eqno(27)$$
The third-body terms cancel, because they are conservative and do no work
around a Keplerian orbit at our level of approximation: 
only tidal-friction terms affect $P$ (or $a$). We can differentiate 
equations (26) w.r.t. time, keeping
$\Jhat$ constant and using equation (22) for any of $\ehat,\qhat,\hhat$:
$$\alpdot\equ -\thin{\Jhat.\thin\K\wedge\hhat\over\vert\Jhat\wedge
\hhat\vert}\w,\w\betdot\equ{\Jhat.\thin\hhat\thin\thin\Jhat.\thin\K-
\K.\hhat\over\vert\Jhat\wedge\hhat\vert^2}\w,\eqno(28)$$
$\K$ being the rotation rate of the frame as given in equation (23). 
The second relation involved some elementary vector manipulations,
but the first came directly from equations (22) and (26), using
$\sin\aJ=\vert\Jhat\wedge\hhat\vert$.

We wish to emphasise the following point, which we believe is treated
incorrectly in much of the literature which we have read.
The rate of rotation of the line of apses, $\olpdot\eq-\betdot$,
is usually attributed to $Z$, the $\hhat$ component of $\K$, to the extent 
that $Z$ is normally referred to as `apsidal motion'. But it is easy to see 
that $\betdot$ in equation (28) can be non-zero on account of the $\ehat,
\thin\qhat$ components $X,\thin Y$ of $\K$ as well, as will happen with a 
massive rotating star whose spin axis is highly inclined to the orbital 
axis and precessing about it. We can see that if $X,Y=0$, so that 
$\K=Z\hhat$, then $\olpdot\eq-\betdot=Z$ as expected. But if $X,Y$ 
are not zero (precession) they contribute to $\betdot$, even in the case
that $Z=0$. Note that this effect does not depend in any way on the details
of our model: it only depends on the fact that the orbital frame $\e,\thin\q,
\thin\h$ has some general angular velocity $\K=X\ehat+Y\qhat+Z\hhat$, while
the system is viewed from a fixed direction $\Jhat$ with variable colatitude 
$\aJ$ and longitude $\bJ$ in the $\e,\thin\q,\thin\h$ frame.

The effect of precession on $\betdot$ is mainly to swing the line of apses back and
forwards (libration), rather than to advance it monotonically (circulation),
as does $Z$. But for the case where (a) $\vO$ is not parallel to $\hhat$, 
(b) the tidal friction terms in equations
(10) and (11) are negligible -- which they usually are -- and (c) the
$\vO$-dependent term in equation (12) dominates over the remaining term --
which is usually the case for rapidly-rotating components -- the librating
and circulating effects are comparable.

Our model can be used to provide times (or phases) of eclipses, when the 
inclination is sufficient for eclipses to occur. We use the simplest 
approximation, that the stars are spherical. For the beginning and end of an 
eclipse we have to satisfy the equation
$$\vert\Jhat\wedge\bd\vert\equ R_1+R_2\w.\eqno(29)$$
Here $\bd$, the vectorial separation of the two stars, has components in
the $\ehat,\qhat,\hhat$ frame given by the usual formula
$$\bd\equ {l\over 1+e\cos\theta}\thin(\ehat\cos\theta+\qhat\sin\theta)\w,
\eqno(30)$$
with $l=a(1-e^2)=h^2/GM$ being the semi-latus-rectum. Equations (29) and (30)
give a quartic equation for $\cos\theta$,
which we solve analytically. The coefficients of the quartic are determined by
$\Jhat.\thin\ehat$ and $\Jhat.\thin\qhat$, which vary with time as the basis 
set moves. Having determined the four, two or zero real roots that lie in the 
range [-1,1], we can determine the phase $\Phi$ (i.e. time, divided by period) 
of ingress and egress from the usual formulae
$$2\pi\Phi=\psi-\sin\psi\w,\w\cos\psi\equ {e+\cos\theta\over 1+e\cos\theta}\w.
\eqno(31)$$
This is phase measured from periastron ($\theta=0$). We can also determine the 
phases of other significant points on the orbit. The point where the projection
on to the orbital plane of the the line-of-sight vector $\Jhat$ intersects the 
orbit (conjunction) is given by $\theta=\theta_1$ say, where
$$(\Jhat\thin-\thin\Jhat.\thin\hhat\thin\hhat)\wedge\bd\equ 0,\w i.e.
\w\tan\theta_1={\Jhat.\qhat\over\Jhat.\ehat}\w.\eqno (32)$$
The point where the radial
component of velocity (relative to the CG) vanishes is given by $\theta=\theta_2$
say, where
$$\Jhat.\dbfdot=0,\w\dbfdot={h\over l}\thin(-\ehat\sin\theta+
\{e+\cos\theta\}\qhat)\w,\eqno(33)$$
so that
$$\sin(\theta_2-\theta_1)\equ e\sin\theta_1\w.\eqno(34)$$
Thus the values of $\theta$ at both these points are also functions of 
$\Jhat.\thin\ehat$ and $\Jhat.\thin\qhat$. There are two values of both $\theta_1$ 
and $\theta_2$ in the range $0-360\deg$; in \S4 we take $\theta_1$ such that $*1$ 
is in front, and $\theta_2$ such that $*1$ is behind. For triple systems we ignore 
the small effect due to the variable motion of the CG of the inner pair.
The points where the radial velocity is a maximum (or minimum) are given
by $\Jhat.\bd=0$ (since $\dbfddot\parallel\bd$ in the unperturbed orbit),
and hence by $\theta_3=\theta_1\pm 180\deg$. 

The model we present here has, we believe, the merit of considerable simplicity,
both conceptually and numerically. We emphasise here the approximations on which 
it is based:
\sep (i) Only the quadrupolar component of the distortion of each star is 
modelled. This assumption may be fairly good in systems which
are only mildly eccentric, but can be expected to be less valid in systems
of high eccentricity. For the distortion due to rotation, it is assumed that 
the stars are in solid-body rotation.
\sep (ii) The components are assumed to adjust instantaneously to fill an
equipotential of the joint gravitational-centrifugal potential. This leads
to a specific tidal velocity field within each star, whose shear, combined
with a viscosity assumed to be due to convectively-driven turbulence, determines
the force of tidal friction. Although some analyses have argued that the
effect of convection in a star with a radiative envelope and a convective core
is small, we follow the analysis of EKH98 which shows that tidal dissipation
within a convective core is not small.
\sep (iii) The effect of the third body is only modelled at the quadrupole
level of approximation. This is sufficient to demonstrate such phenomena as
Kozai cycles (Kozai 1962), where the third body, if placed in an orbit highly inclined to
that of the first two, causes large fluctuations in eccentricity on a long
timescale. The approximation is not very good for a {\it parallel} orbit,
since all the off-diagonal components of the tensor $S$ vanish so that the
only effect in equations (18) -- (21) is the apsidal-motion term. It therefore
does not allow us to model the fluctuations in eccentricity that the third
body produces in the inner pair; actually these are quite small, according
to an N-body integration, but they can be significant on a long timescale
when combined with tidal friction (KEM98).
\sep (iv) The same level of approximation means that there is no additional
force, or couple, on the outer binary. Consequently angular momentum is not 
conserved: the inner binary can gain or lose angular momentum, but not the
outer. Moderate accuracy relies on the fact that the angular momentum of the
outer binary is large compared with the inner, so that a substantial amount
lost by the inner counts as only a small perturbation to the outer. However
angular momentum increases with only the cube root of the period, at given
masses, so that a period ratio of 50 (an unusually small value, but appropriate
to SS Lac) means an angular momentum ratio that is not large.

\section{Some illustrative examples}

Our model for perturbed orbits is original to the extent that (a) it has
a specific formulation for the parallelisation of stellar spin which is 
initially oblique to, or even
anti-parallel to, orbital angular momentum, and (b) it includes the effect
of a third body along with the other perturbations. S{\"o}derhjelm (1975)
gave a formulation of the third-body effect, but without the other effects. 
As applied to systems which are binary rather than triple, and where the 
stellar spins are (at least by hypothesis) 
parallel or nearly parallel to the orbit, our model does not differ from
lowest-order standard analyses. We confirm the following standard results: 
\sep (i) on a timescale
of $\ts t_{\rm TF} \thin I\Omega/\mu h$, the spin becomes parallel to the orbit 
and `pseudo-synchronous', i.e. it reaches the value where the viscous couple $W$ 
in equation (6) is close to zero (Hut 1981). The couple is an average around 
a Keplerian orbit, and it vanishes when the larger but short-lived couple near 
periastron is balanced by the weaker but longer-lived (and opposed) couple at 
apastron. Equating $W$ to zero gives the pseudo-synchronous
value of $\Oh$ as $\omega$ multiplied by a function of $e$.
\sep (ii) On the longer timescale $t_{\rm TF}$ the orbit is circularised.
\pn However, both these statements have to be qualified by the condition that
the spin angular momenta of the stars have to be suitably small when compared
with the orbital angular momentum; otherwise the binary can become
either desynchronised or decircularised.
\sep (iii) For triples, ignoring the effects of quadrupolar distortion, tidal 
friction and GR, we obtain equations which allow
the eccentricity and the mutual inclination of the inner orbit to fluctuate 
periodically between limits (Kozai cycles: Kozai 1962, Mazeh \& Shaham 1979; 
KEM98). If we start with 
$e=0$ and $\sin\aH\tg\sqrt{2/5}$ ($\aH\tgs 39\deg$), these cycles can have 
large amplitude. The maximum eccentricity reached is
$$ e_{\rm max}^2\equ {\scr {5\over 3}}\sin^2\aH-{\scr {2\over 3}}\ \approx 1-
{\scr {5\over 3}}\delta\aH^2\ {\rm if}\ \delta\aH\eq{\pi\over2}-\aH.\eqno(35)$$
We see that $e_{\rm max}$ can approach very close to unity if $\aH$ is only
moderately close to $90\deg$.
The timescale of these cycles is of order $1/C$ -- equation (15) -- and so
of order $P_{\rm out}^2/P$, apart from a mass-ratio-dependent factor which
is only significant if the third body is much less massive than the other two.

   Commonly, among observed binaries, either the orbital period is sufficiently short
that tidal friction has already circularised it, or sufficiently long that
tidal friction is insignificant on a nuclear timescale. This is because of
the high power of $R_1/a$ in equation (7). There is only a fairly narrow
range of periods, perhaps $4 - 5\thin$d (but depending on mass and age) where one
might hope to find binaries in which parallelisation is still taking place.
However, there exists the interesting SMC radio-pulsar binary \smc\ 
(Kaspi \etal 1994) in which it is
conjectured that, as a result of an asymmetric supernova explosion (SNEX),
the neutron star is on an inclined orbit, relative to the spin of the normal
B1V component. In our first example below, we endeavour to model the process 
of parallelisation etc. of the B star in this system. In our second example, 
on the supposition that an SNEX may typically put
a neutron star (NS) into a non-synchronous orbit, we also consider a model
of a massive star with an NS companion, and various degrees of asynchronism.
Such models can experience desynchronisation and/or decircularisation.

There is only a rather small number of known triples where the inclination 
of the outer to the inner orbit is directly measured, and even fewer in which it
is clearly established that this inclination is large enough to cause Kozai 
cycles. In fact, the system $\beta$ Per (Lestrade \etal 1993) is the only
example we know. We therefore consider how the orbit of this system might have 
been modified at an early stage in its life, when it was a detached near-ZAMS
system.

In binary orbits that are eccentric, but already (at least by hypothesis) 
parallelised and pseudo-synchronised, the only part of our model to be testable 
is the effect of tidal distortion and GR on apsidal 
motion. In this respect our model is no different from the classical model:
Claret \& Gim{\'e}nez (1993) have discussed apsidal motion, comparing observed values
with those expected theoretically from the combination of quadrupole distortion
and GR. For many systems there is reasonably good agreement. For some systems 
however there is disagreement, even strong disagreement, for example DI Her 
(Guinan \etal 1994) and V541~Cyg (Lacy 1998). We suspect that the discrepancies
here may be due to the presence of a third body, so far undetected; although
another possible reason for aberrant apsidal motion is that the stars are
rotating obliquely to the orbit. The latter leads to smaller apsidal motion than
expected for parallel synchronism, and the apsidal motion can even have the
opposite sign -- equation (12) -- if $\Oah\tls\Omega_1/\sqrt{3}$. However a third body can
also contribute apsidal motion of either sign.
\subsection{ (i) Parallelisation, synchronisation and circularisation in a wide eccentric orbit.}
We consider the effect of the perturbative forces within a binary
roughly based on the radio-pulsar binary \smc\  in the SMC (Kaspi \etal 
1994, Bell \etal 1995). In this binary (no third body is detected, or
suspected) the pulsar's orbit is quite wide
($P= 51.17\thin$d), and highly eccentric ($e=0.808$). The directly-measured 
longitude of periastron gives $\bJ\eq 270\deg-\olp\equ 154.76\deg$, from equation (25). 
The companion B1V star appears to be unusually {\it in}active: it is not a
Be star, apparently has negligible wind, and the pulsar is not accreting
significantly, at least not enough to be an XR source. These unusual
circumstances (for massive neutron-star binaries) mean that the radio orbit 
is unusually well-defined, so that even the slight change of {\it orbital} 
parameters, due presumably to tidal friction, apsidal motion and precession, 
are measurable. 

We refer to the NS component as $*1$, because it is descended from what was 
presumably the originally more massive component, and the B1V star as $*2$.
This choice determines which suffix belongs to which star.

Although the mass-function of the pulsar is accurately known, there is
only a very tentative radial-velocity curve for the B star. Bell \etal (1995),
assuming $M_1=1.4\Msun$, suggest the following parameters: $M_2\ts 8.8\Msun,$ 
$ R_2\ts 6.4\Rsun,$  $L_2\ts 1.2\times 10^4\Lsun$, with substantial uncertainties.
The consequential inclination of the orbit to the line of sight is 
$\aJ\ts 44\deg$ (or $136\deg$). Bell \etal also estimate a projected rotational velocity 
for the B1V star of $\vr\sin i\eq R_2\vert\vO_2\wedge\Jhat\vert\ets 113\thin$km/s, 
which suggests a rotational period for the star of less than three days, 
but depending on the unknown orientation of the stellar spin relative to the 
observer. The spin rate though not clearly known is marginally consistent
with the possibility that the B star is in pseudo-synchronism (Hut 1981). 
For $e\ts 0.8$ 
pseudo-synchronism requires $\Obh\ets 12.5\omega$ or $P\ts 4\thin$d. It is
very likely however that the NS was put into its current highly eccentric 
orbit by a supernova `kick', which also makes it likely that the stellar 
spin is inclined, perhaps quite highly inclined, to the orbit. It is in
fact easier to account for the rate of orbital period change if the spin
is retrograde, since this tends to maximise the contributions of $W_2$ and
$V_2$ in equation (26).

Kaspi \etal (1996) further determined various rates of change:
$\Pdot/P\ts -2.2\times 10^{-6}\thin$/yr,
$\alpdot=2.1\times 10^{-4}\thin$rad/yr, 
$\betdot=-\thin 4.5\times 10^{-4}\thin$ rad/yr.
The sign of $\alpdot$ will be different if we adopt $\aJ\ts 136\deg$
instead of $44\deg$. The accuracy of these quantities, and $\vr\sin i$, is
of the order of $3 - 10\%$.

Our model is slightly over-constrained by the current observational data, 
supposing that we take literally the estimate (9) for the viscous timescale.
We adopt the values of $M_2,R_2,L_2,M_1,P,e,\aJ,\bJ$ mentioned above.
There is no evidence for a third body, and so we take $M_3=0$. We ignore all 
parameters relating to the neutron star except its mass, since its
spin angular momentum will be too small to influence the system. This
only leaves the three components of $\vO_2$ to be assigned at $t=0$,
and there are four constraints to be satisfied: $\Pdot/P$, $\vr\sin i$,
$\alpdot$ and $\betdot$ should all have the values listed above. 

Fig 1 is a short evolutionary run starting from $\Omega/\omega=20$, 
$\aOb=135\deg$, $\bOb=0\deg$. On such a short timescale only $\bOb$ 
changes significantly, by precession. The plotted quantities are the 
ratios of the computed to observed values for the four quantities 
listed. It can be seen that at 195$\thin$yr all four ratios are fairly 
close to unity, at which point  $\bOb=111\deg$. We therefore adopt this 
as a new starting value.

\centerline{FIG 1}

Since the starting values used above were a shot in the dark, we can
expect to get better agreement by some procedure such as least-squares.
However, the answers will be very strongly dependent on (a) the radius
$R_2$, which enters to the eighth power in equation (7), and (b) the
estimate (9) for $\gamma$, which must be very uncertain. It might be
more realistic to treat $\gamma$ as an unknown, in which case probably
an exact solution (and possibly several, because of the non-linearity
of the equations) can be found; but it will still be very dependent on 
$R_2$, which cannot be accurately known. Thus we feel it is premature
to attempt a definitive solution, but we feel encouraged by the fact
that the model is not obviously wrong.

The evolution of the eccentricity, of the component of spin parallel to the 
orbit (relative to the total spin), and of the orbital period (relative to 
initial period) 
is shown in Fig 2a, for a timespan of about $3\thin$Myr into the future. 
We started with the parameters listed above (but $\bOb=111\deg$).
The evolution of the B star in this interval has not been allowed for: 
the main-sequence lifetime of an $8.8\Msun$ star is expected to be about 
$33\thin$Myr. The perpendicular spin goes through zero at about 
$0.5\thin$Myr, and the spin is almost completely parallel by
$1.7\thin$Myr. Circularisation takes a good deal longer, and is only
half-complete by $3\thin$Myr -- but it will start to be strongly accelerated
by the neglected evolutionary expansion, at this stage. Currently 
$\edot/e\ts-2.5\times 10^{-7}$/yr, on our model. This is comfortably below 
the upper limit found by Kaspi \etal (1996) of $7\times 10^{-6}$/yr.

\centerline{FIG 2}

Fig 2b shows the two components of spin in the orbital plane, $\vO_2.\ehat$ and 
$\vO_2.\qhat$, plotted against each other. To make the figure clearer
the viscous evolution was speeded up by a factor of $10^{8/3}$; this makes
for a much less tight spiral pattern. Evolution starts slightly left-of-centre at the top edge. 
The rotation axis precesses counter-clockwise about 1.25 times, until 
the vertical component of $\vO_2$ ($\vO_2.\hhat$, Fig 2a) changes 
sign, and then precesses clockwise while the two horizontal spin-components 
diminish towards zero. Had we kept to the more realistic viscous timescale
of Fig 2a, there would have beeen about 550 revolutions of the axis before
it reversed direction.

Fig 2c shows the evolution of four timescales, also using the speeded-up
model of Fig 2b. The timescales are all given as logs, and in years. The timescale
for period change $\vert P/\Pdot\vert$ (plusses) starts at $\ts 10^{3.1}\thin$yr, which
would be roughly the required value of $\ts 5\times 10^5\thin$yr if we did not 
speed up the viscous evolution by
$10^{8/3}$. The eccentricity timescale $\vert e/\edot\vert$ (asterisks) is about 6 times
longer to start with, but is more nearly constant. The two other timescales 
shown are both related to apsidal motion: $1/Z$ (circles), and $1/\vert\betdot\vert$
(equation 27; crosses). $\betdot$ is the actual apsidal motion, which however is 
influenced by the precessional terms $X,\thin Y$ as well as by the usual
`apsidal motion' term $Z$. Prior to about $1000\thin$yr, when the vertical
spin passes through zero in the speeded-up model, the line of apses
turns at a highly variable rate; probably the axis was librating rather than 
circulating, in the fairly recent past. Once the B star is no longer
counter-rotating the line of apses circulates more uniformly, but with an 
oscillating component which diminishes as the spin becomes
more parallelised.

Figure 2d shows two more timescales (also logged): the precessional timescale, 
$1/(X^2+Y^2)^{-1/2}$ (plusses),
and the the timescale of the change of inclination of the orbit to the 
line of sight, $1/\vert\alpdot\vert$ (asterisks). The precessional rate
goes through zero once, causing the cusp at $\ts 1000\thin$yr. The
orbital direction oscillates about zero, causing many cusps in the log 
modulus of its derivative.

The origin of the present system presents some puzzles, and has been the subject
of recent controversy (van den Heuvel \& van Paradijs 1997; Iben \& Tutukov 1998;
hereinafter HP, IT). HP favour a history that involved Roche-Lobe overflow (RLOF)
followed by a supernova explosion (SNEX) with an asymmetric kick, and IT a history that involved
a common-envelope (CE) phase followed by an SNEX without a kick. We believe that
neither history is satisfactory, and propose a scenario which is somewhat similar
to IT in its earlier phase (but requiring a less massive progenitor to the NS),
and rather like HP in its later phase, requiring an SNEX kick.

   We would normally expect that the system, having started with
two massive MS stars, would have evolved through RLOF, so that $*1$ (the
{\it originally} more massive component) would have become a helium star,
perhaps with a modest H-rich envelope, before heading on to C-burning
and so fairly quickly to an SNEX (HP). However two things argue against this:
\sep (i) If $*1$ was originally over $\ts 10\Msun$, enough to leave a post-RLOF remnant 
capable of an SNEX, then RLOF should have made $*2$ considerably more
massive than it is now (even though its mass is by no means certain). 
In such RLOF we normally expect $*2$ to become more massive than the
{\it original} mass of $*1$.
\sep (ii) We would expect that as a second result of the RLOF $*2$ would be a very 
rapid rotator, a Be star more-or-less, instead of the rather slowly rotating and 
unusually inactive star observed. 

A possible answer to both these points is that 
\sep (a) the initial $*1$ was only moderately more massive than $*2$ now, say $\tgs 12\Msun$
\sep (b) the initial $*2$ was little different from the $*2$ now seen (the B1 
star) 
\sep (c) the binary was fairly wide initially, say $P\tgs 50\thin$d
\sep (d) $*1$ evolved to a point where its outer layers, helped by the 
disturbing effect of the binary companion, became unstable and blew away, firstly 
as a P Cyg star and then as a Wolf-Rayet star, perhaps without $*1$ ever reaching 
a radius as large as its Roche-lobe radius. 
\par\noindent If $*1$ did reach RLOF, this might have been more like a CE event, with
much of the envelope disappearing to infinity rather rapidly, and with only moderate, 
or perhaps even negligible, orbital shrinkage. But we would rather categorise the
process as `binary-enhaced stellar wind' (BESW), which may have altogether prevented
$*1$ from ever reaching its Roche lobe. 

The WR binary $\gamma^2$ Vel, with $P\equ 78.5\thin$d and $e\equ 0.33$
(Schmutz \etal 1997) might be of the same character as the possible 
immediate precursor to the \smc\ binary, 
since the high eccentricity argues against there having been an episode of
Roche-lobe overflow. The $*2$ of $\gamma^2$ Vel is an O8III star of $20 - 30
\Msun$, substantially more massive than we require. Consequently $*1$ would
also have been substantially more massive originally, perhaps by about the same factor.

IT's model was somewhat similar to ours, except that they argued for a more massive initial 
$*1$, $\ts 28\Msun$. This was required because of their desire to produce the
configuration of \smc\ {\it without} an SN kick. They argued for a CE event
which reduced the period from an initial value of $27 - 76\thin$d to a value
of $\ts 3.2\thin$d. They 
postulate that the obliquity of the spin to the orbit, strongly suggested by
the measured $\alpdot$ of Kaspi \etal (1996), is simply left over from a
primordial obliquity, and survived any possible tidal friction during the
helium-star phase, when in their model the orbital period was 3.2$\thin$d.
A difficulty with this is that with $*1$ initially so much more massive than
$*2$, $*2$ should be rather little evolved, and should be substantially
smaller than the value of $6.4\Rsun$ suggested by Bell \etal (1995). Our
model supposes a much less massive $*1$, and so allows $*2$ to be more
substantially evolved. Our model does not predict, nor need to predict,
the orbital period during the helium-burning phase; we accept the probability
of an asymmetric kick, which could in principle lead to the present period
if the intermediate period was anywhere in the range of $\ts 3 - 50\thin$d.

Our model of the current orbital evolution might give an upper limit to the 
age of the system (since the SNEX),
by integrating backwards from present conditions. This is not a very safe
process, numerically, in a dissipative system, but we made an estimate of the 
accuracy by integrating forwards again. It appears that in fact the evolution
{\it decelerates} going backwards, as is hinted at by the behaviour of
$\Obh/\Omega_2$ in Fig 2a. We integrated back $\ts 9\times 10^5\thin$yr, reaching 
$P=200\thin$d, $e=0.922$ and $\aOb=149\deg$; on integrating forwards 
again we recovered $P$, $e$, $\aOb$
and $\Omega_2/\omega$ to 5 significant figures, while $\bOb$ was in error
by about $\ts 80\deg$ after several thousand rotations of the $\vO_2$ axis.
The spin period $\ts 9\times 10^5\thin$yr ago is predicted to have been $1.6\thin$d.
Although it is marginal, this may be consistent with the B1V component's
still being reasonably inactive, so that the model is still applicable.
Thus it is possible that the system may be as much as $\ts 10^6\thin$yr old
in its present form. The required orbit so long ago might seem improbably
long and eccentric; but one might reasonably think the present orbit
improbably long and eccentric if it had been hypothesised rather than
measured.

The \et model of \tf has often been considered inadequate for systems like
\smc. There appear to be two main reasons, one of which we largely accept
and the other we reject. In order, they are
\sep (a) near-equilibrium is not very likely to be established in a highly
eccentric orbit; it is more reasonable in a nearly circular orbit (like
the Earth-Moon system)
\sep (b) Although turbulent convection may be a good source of friction
in stars with deep convective envelopes, massive stars are only convective
in their cores where the amplitude of the tide is considered to be too
small to be significant. Radiative damping in the outer layers might
contribute, but this is orders of magnitude smaller.
\pn We believe that (b) is largely based on a highly inexact estimate of the \et 
velocity field.

If the principle is accepted that surfaces of constant density (and pressure)
are always closely equal to equipotential surfaces (the basic assumption
of the \et model) then presumably the velocity field is determinate, and
comes basically from conservation. Alexander (1973) and Zahn (1977, 1978) made
crude estimates, based on the motion being assumed either incompressible or
irrotational, and concluded that the amplitude of the tide (whose square is 
proportional to the rate of dissipation) goes to zero like $r^4$, approaching
the centre. If the convective core were, say, one third of the stellar
radius, then the dissipation would be down by $\ts 10^{-4}$ relative to a
star with a largely convective envelope. However, EKH98 determined -- their
equations (100) to (112) -- an expression for the tidal velocity field, and
its rate of viscous dissipation, which is {\it exact}, to the extent that 
(a) the \et model is exact, and (b) dissipation
is primarily by the effective viscosity of turbulent eddies. The velocity
field is neither irrotational nor incompressible, nor does it diminish
to zero like $r^4$. Rather, the tidal amplitude diminishes from its surface 
value by less than a factor of 10 for typical MS models. Thus the effect
of dissipation in the convective core is by no means negligible: it may be
down by $\ts 10^{-2}$ only. This is the basis for our estimate of $\gamma$
in equation (9). In the Appendix, we briefly summarise the analysis of
EKH98 regarding the factor $\gamma$.

Witte \& Savonije (1999) computed the spectrum, and damping rates, of normal 
modes that can be expected to be excited in a $10\Msun$ star as a result of 
perturbation by an NS companion with the orbital parameters of \smc. They 
obtained a braking timescale that was usually in the range $10^{5.5} - 
10^{6.5}\thin$yr. The timescale changes rapidly by factors of $\ts 10$, both 
up and down, on timescales of only $10^4\thin$yr or less. There are occasional 
excursions to values of braking timescale as low as $10^3\thin$yr, which result 
from modes resonating with harmonics of the orbital frequency. There are also 
occasional episodes of orbital spin-up rather than spin-down. Such a detailed 
model may well be demanded by the physics,
but inevitably means that the interior structure of the star will have to
be very precisely known: a good deal more precisely than information which is
currently available. We hope that our estimate, equation (7), can serve as
a crude average over a range of time of more detailed values that can only
be computed if the structure and rotation of the star are known to
considerable accuracy.
\subsection{(ii) The Darwin and eccentricity instabilities}
Inherent in equations (1) - (4) are at least two kinds of instability. Firstly,
there is the Darwin instability. Consider the case of a binary (i.e. no third
body) where the spin of one massive component ($*2$) is parallel to the orbit, and the
companion ($*1$) is a point-mass neutron star, as in the previous example. Equation (25)
for $\omegadot$ can be united with equation (4) for $\Omegadot_2$ to give
$$t_{\rm TF2}\thin{d\over dt}\log{\Omega_2\over\omega}\equ {\mu h\over I_2\Omega_2}\thin W_2
-\left(W_2+{e^2V_2\over1-e^2}\right)\hskip 1truein$$
$$\hskip 0.35truein=\left[{\mu h\over I_2\Omega_2}f_a(e)-3f_b(e)\right]-
{\Omega_2\over\omega}\left[{\mu h\over I_2\Omega_2}f_c(e)-3f_d(e)\right]\ ,\eqno(36)$$
where the functions $f_a, \dots, f_d$ are all functions of $e$ that can be evaluated
from equations (5) and (6). All these functions tend to unity as $e\to 0$. It can be 
seen that as long as
$${\mu h\over I_2\Omega_2}\ \tg\  {3f_d(e)\over f_c(e)}\ets 3\w{\rm if}\w e\ets 0\w,
\eqno(37)$$
then $\Omega_2\to\omega$ as time increases. But if inequality (37) is violated,
$\Omega_2/\omega$ will diverge as time increases. For a value of $e$ which is
not small, there still is a critical condition but it is $e$-dependent. This 
well-known instability reqires that the spin angular momentum $I_2\Omega_2$ must be 
greater than a third of the orbital angular momentum $\mu h$ (for $e=0$).

The eccentricity instability is seen in equations (18) and (5). Specialising
once again to the situation where one star is a point mass, and $e\equ 0$, we see
that if
$$\Obh\tg {18\over 11}\thin\omega\w,\eqno(38)$$
then the eccentricity starts growing exponentially. If $e\tg 0$ to start start 
with, there is still the possibility of $e$ growing, although the criterion is 
now $e$-dependent. In other words, if the star is rotating fast enough, it gives
up its angular momentum in spurts sufficiently concentrated towards periastron
that the companion star is flung into a wider and wider orbit -- but with
periastron not much changed because that is where the largely tangential impulse
peaks.

\centerline{FIG 3}

Although both instabilities give exponential growth the result can sometimes be
surprisingly self-limiting. Fig 3a shows the evolution of a system whose initial
configuration was unstable to both processes. We took a very massive star
($40\Msun$) evolved substantially across the MS (to $20\Rsun$), put it in 
a $6\thin$d, $e=0.1$, orbit with a neutron star of $1.4\Msun$, and started it in parallel 
rotation at twice the orbital rate. We used the default values (8) of
moment of inertia and quadrupolar distortion. Both the eccentricity and the degree 
of non-corotation (measured by
$\Omega/\omega$) began to grow. For stars of comparable mass it is difficult to
violate criterion (37), but if one star is much more massive than the other,
it can also be large enough, without quite filling its Roche lobe, to be
Darwin-unstable. However although the star spins up relative to the orbit, the
orbit gains angular momentum, and so loses angular velocity, fast enough for
the D-stable criterion (37) to become satisfied later. After between $10^6$ and
$10^7\thin$yrs, the orbit first becomes D-stable and later E-stable, and tends
to both synchronism and circularity with a period of $\ts 30\thin$d. However,
as before we have ignored nuclear evolution in the massive component, which
would no doubt fill its Roche lobe in little more than $10^6\thin$yr.

Fig 3b is the same system except that the stellar spin rate was started at 70\% of
corotation, rather than twice. This is substantially E-stable and very slightly
D-stable to start with, but as $e$ decreases towards zero, and $\Omega/\omega$ 
increases (though only very slightly) towards unity, at about $5000\thin$yrs 
the system crosses the D-unstable margin. Although both
$\Omega$ and $\omega$ are going up, trying to approach synchronism, while $I\Omega$
obviously goes up $\mu h$ goes {\it down} because the orbit shrinks. Hence the
D-stable criterion (37) crosses into instability and the system begins to move rapidly 
away from corotation. The orbit continues to circularise, but desynchronises and
shrinks rapidly towards a collision at $\ts 19000\thin$yr.

\subsection{(iii) Kozai cycles with tidal friction.}
We now consider a problem which has a third body as well as quadrupole
distortion and tidal friction.
When the outer binary is sufficiently inclined to the inner binary  it is 
possible for the eccentricity of the inner binary to fluctuate slowly by a 
large amount (Kozai cycles). The amplitude of the eccentricity fluctuation depends only on the
inclination, and not on the outer period, or eccentricity, or third-body mass;
the period of the fluctuation is of order $2\pi\sqrt{1-e^2}/C$ -- equation (15).
Even if the inner binary, when it is nearly circular, is too wide for 
tidal friction to play a role, the increase in eccentricity may make tidal 
friction important at some point in the Kozai cycle. Recall that $a$, like 
$\omega$ and $\thin P$, is unaffected by the third body at our level of approximation, 
as shown by equation (27), so that as $e$ increases the periastron separation 
decreases. We illustrate this with the well-known semidetached binary Algol 
($\beta$ Per), which has a third body$(\ts 1.7\Msun$) in a $679\thin$d orbit
inclined at $100\deg$ to the semidetached pair's orbit (Lestrade \etal 1993).

In its present configuration, the inner pair is {\it not} subject to Kozai 
cycles, because the perturbation due to the quadrupole distortion of the
lobe-filling component is much larger than the perturbation due to the third 
body. However, at an early stage in its life $\beta$ Per must have been a
detached binary of two near-ZAMS stars, with radii and therefore
quadrupole moments substantially smaller than at present.

If we believe that $\beta$ Per has evolved without mass loss (ML) or angular
momentum loss (AML), i.e. conservatively, we would be able to infer the
period at any mass ratio, from
$$P\ \propto\ {(1+q)^6\over q^3}\w,\w q\eq{M_1\over M_2}\w.\eqno(39)$$
Taking an illustrative $q_0=1.25$, the present period $P=2.87\thin$d and
$q=0.216$ imply that $P_0\ts 0.6\thin$d. However, although we accept
provisionally that ML may have been negligible, there is direct and indirect
evidence that cool Algols experience AML, presumably by magnetic braking in a
stellar wind (Refsdal, Roth \& Weigert 1974, Eggleton 2000a). 
For given masses, the period goes like $h^3$, and so if the
system lost 50\% of its angular momentum, it must have started with $P_0\ts
4.8\thin$d.

What we show in this subsection is that the initial period, had it been longer
than $\ts 3\thin$d, would have shrunk, by a combination of Kozai cycles and 
tidal friction, to a value under $3\thin$d in a fairly short interval of time
$(\tls 10^7\thin$yr). Consequently we have an upper limit to the amount of AML
that could have taken place subsequently, once $*1$ became a cool subgiant
subject to magnetic braking (Eggleton 2000b): about 40\% of the initial angular momentum.

Fig 4 shows the evolution of a `proto-Algol' system with an initial period of
$5\thin$d, in (a) the short term, (b) the medium term, and (c) the long term.
The initial Kozai cycles reach up to $e=0.67$ (starting somewhat arbitrarily
at $e=0.1$). This value is well short of the maximum that would be reached
($e=0.985$) if quadrupolar distortion was negligible, but is nevertheless
quite large. Tidal friction near periastron at the peak of the Kozai cycles
reduces the range of variation of $e$, though somewhat unexpectedly by
increasing the minimum eccentricity even more than by reducing the maximum. By
about $10^6\thin$yr the eccentricity fluctuates between $0.47$ and $0.53$,
and both the range and the mean reduce until by $10^7\thin$yr the orbit
is circularised at $P\ts 2.1\thin$d.

\centerline{FIG 4}

A point to note is that the inclination $\aH$ of the inner orbit to the outer
orbit changes somewhat during the process. We started from $97.5\deg$,
in order to end up with the currently observed value of $100\deg$. For longer
initial periods the change is larger, which probably means that the period
was not in practice much larger than $\ts 10\thin$d before the Kozai cycling
and tidal friction reduced the period to $\ts 2\thin$d.

It is not clear how triple systems, and especially such close triple systems,
formed in the first place, but a possible mechanism, arguably the least
unlikely, is that, fairly early on in the star-forming process when
the stellar density was higher than it is now, two primordial binaries 
had a near-collision, with one component of one binary captured by the other 
binary, and the other component ejected. In this scenario, angles near 
$90\deg$ are much {\it more} likely than those near $0\deg$.

Table 1 shows how the the period $P_{\rm end}$ at the end of the shrinkage process depends
on the period $P_0$ at the beginning, for our specific proto-Algol system. It also shows 
the time taken in the shrinkage and circularisation process, which is always small
compared with the expected nuclear lifetime of the system ($\ts 1\thin$Gyr), and
gives the starting value of mutual orbital inclination $\aH$ that will end up as
the current value of $100\deg$. Assuming that this inclination is distributed randomly,
in a capture process, the range $80 - 100\deg$ has probability $\ts 17\%$, and the
range $86 - 94\deg$ about $7\%$.

\centerline{TABLE 1}

The combination of Kozai cycles plus tidal friction should mean that there
is a shortage of triple systems with (a) fairly high inclination of one
orbit to the other, and (b) inner periods above perhaps 3 -- 4$\thin$d.
This will be difficult to confirm, because it is very difficult to
determine the inclination of one orbit to another. SS Lac (below) is
a triple in which we infer $\aH\ts 29\deg$, which is not enough to give
significant Kozai cycling; thus the inner period of 14$\thin$d does
not conflict with our conclusion. If inclinations are indeed distributed 
at random, then $\ts 50\%$ of triples have $60\deg\tls\aJ\tls 120\deg$,
and a quite substantial deficit of systems with inner periods longer than
say $3 - 4\thin$d can be expected.

\section{An application to SS Lac}

SS Lac is a binary which eclipsed before about 1950, but not subsequently.
A likely explanation was the presence of a third body, in a non-coplanar
orbit, and this was confirmed by Torres \& Stefanik (2000) -- hereinafter 
TS00 -- who found long-period orbital motion in the CG of the 
short-period pair. By coincidence, the longer period in SS Lac is exactly the same
as that in Algol (679$\thin$d). Following TS00, we refer to the 3 components as Aa 
($*1$), Ab ($*2$) and B ($*3$), and the two binaries as A and AB. TS00 
also re-analysed historic light curves of the period
1890 -- 1930, obtaining a mean light curve assigned to epoch 1912. Their
spectroscopic data refers to epoch 1998. In this Section, we model the
dynamical evolution over the period 1912 -- 1998, trying to find a model
which gives the end of eclipses in 1950.
\par In general, our model requires 25 input parameters, which we list
here in two groups:
$$Q\Aa,\thin I\Aa,\thin \vO\Aa,\thin L\Aa;\w Q\Ab,\thin I\Ab,\thin \vO\Ab,
\thin L\Ab\eqno(40)$$
$$M\Aa,\thin R\Aa,M\Ab,\thin R\Ab;\w P\A,\thin e\A;\hskip 1truein$$
$$\hskip 1truein\w \w M\B,\thin P\AB,\thin  
e\AB;\aH,\thin \bH,\thin \aJ,\thin \bJ\w.\eqno(41)$$
However, the A binary is sufficiently wide ($P\ts 14\thin$d) that, provided
its eccentricity (or more specifically its perihelion separation) does
not vary, perhaps intermittently, by a substantial factor, tidal friction
should be quite unimportant. More helpfully still, the $Q$-dependent
distortion terms that determine $X,Y,Z$ in equations (1) -- (4) are
unimportant compared with the third-body terms (components of the tensor 
$S$) in these equations, so that all of the quantities listed in (40) are negligible. 
The radii $R_i$ and the angles $\aJ,\bJ$ defining the direction to the
observer, are also unimportant for the orbital evolution, although 
they matter for the eclipses, and the date of their cessation. Although the
$Q_i$ have little influence on the orbit, they do have a marked effect
on the spins of the stars, because of the couple they cause -- as we mention
briefly below.

This reduces our significant input file to the 13 quantities listed as (41). 
Of these, $e\A,P\A,e\AB,P\AB$ are
well or very well determined at epoch 1998 (TS00). Although $e\A$ may
have (indeed will have) changed since epoch 1912, the other 3 quantities
can be supposed constant. This is because, in our model, 
\sep(a) at the level of the quadrupole approximation for the perturbing 
force of the third body, the AB orbit is exactly constant, and
\sep(b) the perturbing force on the A orbit, in the absence of tidal
friction, is a potential force and hence does not supply energy to
the A orbit when integrated over an approximately Keplerian orbit;
this means that the semi-major axis $a\A$, and the period $P\A$, are
constant even though $e\A, h\A$ are not. 
\pn Thus among these four quantities only $e_A(1912)$ must guessed -- and 
ultimately solved for on the basis that in 86 years the 1998 value (0.136) 
must be reached (in conjunction with other constraints).

Similarly the 3 masses are constrained by but not uniquely determined
by 3 observed mass-functions at epoch 1998. We need the two 1998
inclinations $i\AB,i\A$. TS00 estimated the latter on the basis that (i)
$i\A(1912)=87.6\deg$ is known from the eclipse analysis, (ii) those data 
implied that $i\A$ must have been $81.6\deg$ in 1950, when eclipses
ceased, and (iii) $i\A$ has been decreasing
at a constant rate since 1912. We find that in general the rate of
change of $i\A$ is not very constant, and so we make a guess at the
1998 value of $i\A$, which has of course to be consistent with the value 
that emerges from the calculation. The starting value $\aJ$ is just
$\aJ\eq i\A(1912)\equ 87.6\deg$.

\centerline{TABLE 2}

We also have to know or guess $i\AB$. This however is constant in time since 
$\Hhat$ -- see (a) above -- and $\Jhat$ are constant vectors
in space, even though their components in the $\e,\q,\h$ frame vary as the
frame itself rotates. Dotting the vector $\Hhat\equ (\sin\aH\cos\bH,
\sin\aH\sin\bH, \cos\aH)$ into the vector $\Jhat\equ (\sin\aJ\cos\bJ,
\sin\aJ\sin\bJ, \cos\aJ)$ we have
$$\cos i\AB\equ \cos\aH\cos\aJ+\sin\aH\sin\aJ\cos(\bH-\bJ)\w.\eqno(42)$$
We therefore have to know or guess $\aH$ and $\bH-\bJ$ in 1912, $\aJ$
being known.

TS00's analysis of the $\ts 1912$ light curve gave an inclination of
$87.6\deg$, as mentioned above. They also obtained the longitude of
periastron $\olp$, related to $\bJ$  by equation (18).
The $\bJ$ from TS00's 1912 light curve is $121.6\deg$ (see their Table
6, giving $\olp$, and differing slightly from their Table 5 value for 
reasons which they
explain). However, this value is based on the assumption that $e\A$
is constant, and we find that generally it is not. The most significant
orbital quantity which is given by the light-curve analysis, as TS00
explain, is the departure $\dfII\equ -0.072$ of eclipse II from 
phase 0.5 relative to eclipse~I. For moderate eccentricities,
$${\pi\over 2}\dfII\ \approx\ e\A\cos\olp\eq-e\A\sin\bJ\equ -0.1128\w.\eqno(43)$$
For $e\A\equ 0.136$ this gives the value of $\bJ$ mentioned above. But
in our best near-solutions we usually find $e\A$ increasing, i.e. it
started in 1912 with a smaller value. Evidently it cannot have been smaller 
than 0.1128. Our preferred starting value is $ 0.115$, and this
implies $\bJ\equ  101.2\deg$. A value above rather than below $90\deg$
is preferred, because TS00's value of $e\A\sin\olp\equ -e\A\cos\bJ$, 
though substantially less well determined, is fairly definitely
positive.

It may seem rather unsatisfactory that our preferred $e\A(1912)=0.115$ is
very close to the minimum value $0.1128$ inferred from eclipses. However,
what can be seen as `special' about the system is rather the fact that
the 1998 value of $\bJ$ is extremely close to $90\deg$: $91.7\pm0.6$ (TS00,
their Table 2, giving $\olp\equ 178.3\deg$). Such a value, viewing the
system almost exactly along the latus rectum, favours the maximum departure 
($\dfII$) of the secondary eclipse from phase 0.5. If we imagine,
going backwards in time, that $e\A$ does {\it not} change, then we are driven to
postulate a rather large change in $\bJ$, to the TS00 value $121.6\deg$, to 
allow for the fact that
the eclipses were substantially closer than this maximum value in 1912. What we
conclude here is that less apsidal motion was necessary, because the
eccentricity was a little smaller in 1912. We would quite generally
expect that eccentricity changes on the same timescale as apsidal motion.
Both timescales are dictated primarily by the coefficient $C$ in 
equation (15), if as for SS Lac only the third-body perturbation is
significant.

Our guess at the initial value of $e\A$ therefore provides us, from eclipse 
data, with a starting value for $\bJ$ via equation (43). We already have 
$\aJ\eq i\A\equ 87.6\deg$ from TS00's light curve data. We have to make two further
guesses, at the angles $\aH$ and $\bH$  which in 1912 gave the orientation
of $\Hhat$.

To summarise, of the 13 quantities we need to start with in 1912,  4 are
known directly from observation: these are $P\A,\thin P\AB,\thin e\AB$ from the 1998 radial
velocity curves and $\aJ\eq i\A$ from the 1912 light curve. If we then guess 
the following 4 quantities -- $i\A$ in 1998, and 3 starting values $e\A,\thin\aH$ 
and $\bH$ in 1912, we can work out the remaining 6 starting values from the 
following 6 observationally determined quantities: 3 mass functions from the 
1998 radial velocity curve, and
2 fractional radii and the phase lag $\dfII$ from the 1912 light curve.
Having integrated the equations for the $86\thin$yr timespan, we then have 
4 further pieces of observational data to constrain our 4 guesses. Three of these
are $e\A$ and $\bJ\eq 270\deg-\olp$ 
in 1998, and the cessation of eclipses in 1950. We determine a theoretical $\Te$, the
time of cessation, as the average of the two times after $t=0$ (1912) at which 
the two series of eclipses (primary, and secondary) stopped. The observational 
value to match is $\Te\equ 38\thin$yr. The fourth and last constraint on 
the four guesses is that the value for $i\A$ in 1998 should equal the
value guessed in the first place. Table 2 lists the values used, 
taken from TS00, and also lists our approximate solution.

Table 2 groups parameters under `observed', `guessed', and `computed'.
All the observational data are taken from TS00. Our `guess' was based on a preliminary
eyeball search of parameter space, and refined by trial-and-error.

Since the differential equations are non-linear, there is no guarantee
either that a solution satisfying all the constraints exists, or that 
if it does it is unique. However our very
brief search located one quite accurate solution with a fairly
modest inclination between the orbits ($29\deg$), and a rather more 
extended search suggested that there were unlikely to be any other solutions, 
except possibly at high inclination where the behaviour can become rather
chaotic. Another possibility, which we have not explored, is that the orbital 
inclination has decreased from $92.4\deg$ rather $87.6\deg$ in 1912.

\centerline{FIG 5}

Figs 5a and 5b illustrate some aspects of the model. Fig 5a
follows the orbital evolution for 1912 -- 1998, and Fig 5b follows it
for slightly over $3000\thin$yr. Date is plotted horizontally, and phase
(from 0 to 2, so that two complete cycles are shown) vertically. In Fig 5a,
eclipses occurred within the long cigar-shaped areas labelled as I 
($*1\eq$Aa eclipsed by $*2\eq$Ab) and II (Ab eclipsed by Aa). Eclipse I 
was slightly deeper and narrower, in 1912 (TS00). Phase in this figure is 
measured from periastron. Also shown as (i) and (ii) are the phases
of points determined by equations (32) and (34). 

In Fig 5b, the same information is given for a much longer timespan: 1912 -- 5250.
Small leaf-shaped patches now indicate the episodes of eclipses, and the curves
indicating the phases (i) and (ii) slope generally downwards because of apsidal 
advance (and other rotation of the orbital frame).
It can be seen that the next series of eclipses is not to be
expected until about 2500. Epochs when eclipses take place appear to be
separated alternately by a long interval and a shorter interval, and
unfortunately we seem to be entering a long interval. Each period of eclipses
lasts about a century. 

The fact that the phases of (i) and (ii) decrease (mostly, but not always)
relative to the phase of periastron, as seen in Fig 5b, is of course due
to the motion of the orbital frame. If this motion were just apsidal
motion, i.e. if the major axis were rotating only about the angular
momentum axis, we would have a relatively simple relation between the
anomalistic period (periastron to periastron) and the sidereal period
(successive passages through a plane fixed in an inertial frame and
containing the CG), and lines (i) and (ii) in Fig 5b would have a constant
slope. But with precession as well, the relation between
the two periods can be rather complex. From the point of view of our simple
model it is the anomalistic period that is `basic': if the perturbing forces,
however many of them, are conservative, the anomalistic period $P$ as given
by equation (16) is a constant at our level of approximation.

The period of precession of the inner orbit is about $1000\thin$yr, and the
inclination to the line of sight ($\aJ$) oscillates between about $47\deg$
and $105\deg$. The inclination of the inner orbit relative to the outer oscillates
by only about $1\deg$. It is inherent in our level of approximation
that the inner angular momentum should be small compared with the outer, and
unfortunately this is hardly true for SS Lac; but the fact that the inner 
orbit oscillates so little may nevertheless make the solution reasonably valid.

We do not discuss the accuracy of the input data, and of our fit, in detail, 
for four reasons:
\sep (a) TS00 discuss fully the accuracy of the observational data. 
We have used only values which are independent of
their assumptions that (i) $e\A$ is constant, and (ii) $\aJ\eq i\A$ and 
$\bJ\eq270\deg-\olp$ change at constant rates. All the 
standard errors are less than 1\%, except for $e\AB$ (13\%), 
$\dfII$ (6\%), $f\B$ (3\%), $R/a$ (3\%) and $\Te$ (3\%); $e\AB$ only 
appears in $C$, equation (15), and in a very non-sensitive way.
\sep (b) We have zero degrees of freedom -- 4 constraints to satisfy,
with 4 unknowns -- and so if we can find a solution at all it will be
exact, to the extent that the data is. A hypothetical problem is that
there might be functional dependences among the constraints, but the 
fact that our eyeball search converged very rapidly suggests there are
not. Varying each of our 3 guessed angles by 1$\deg$ usually gave a much worse 
fit, and so did varying $e\A(1912)$ by 0.001. Hence we believe that the 
guesses are right to about this level of accuracy.
\sep (c) By defining the computed value of $\Te$ as the average of the
two times at which the two series of eclipses disappear, we make it
a discontinuous (stepwise) function of the input, and cannot therefore 
differentiate it smoothly. We could develop a more sophisticated
definition, but this seems unnecessary in view of the rather good
solution found by trial and error.
\sep (d) The main uncertainty is possible systematic error, such
as the possibility that equations (1) -- (4) are wrong. We had hoped
to find more constraints than unknowns, and so test the theory more
rigorously. 

We can however make some predictions which are testable:
for example the inclination $i\A$ should decrease to $70\deg$ in
2011, and to $65\deg$ in 2039. This should produce a measurable change
in the mass-function. The eccentricity should be currently approaching 
its peak of $\ts 0.138$, and so may not change significantly for about $30\thin$yr,
but should drop to 0.132 by 2040. It should reach a minimum of 0.09 in
2160.

\centerline{FIG 6}

Fig 6 illustrates two possible behaviours of the spin $\vO_1$. The 3
components in the instantaneous orbital frame are shown as functions 
of time. The stars were started, arbitrarily, with $\vO=\vecomega$. If the 
stars were perfect spheres they would simply maintain constant (vectorial) 
spin in an inertial frame -- tidal friction being negligible in this system 
-- and so oscillate sinusoidally in the frame of the precessing inner binary. 
But because they have quadrupole moments, due partly to their spin and 
partly to their gravitational effect on each other, there are couples on 
them. In Fig~6a we used our default value of $Q=0.028$ ($n= 3$ polytrope),
and in Fig 6b reduced this to 0.01. We have tried other values of $Q$, and 
do not see any very simple relation between the size of $Q$ and the 
amplitudes or other characteristics of the oscillations. Considering that 
the orbit precesses on a cone of half-angle $29\deg$, it seems surprising 
that the rotation axes of the component stars (we only plot $*1\eq$Aa) can 
turn by more than $90\deg$ in the course of $\ts 500\thin$yr.

It may be questioned whether the approximation that we make in this paper,
that a star rotates with a unique $\vO$ as if it were rigid, is sustainable
in circumstances where $\vO$ is changing in direction by a large amount in a few hundred
years. Tidal friction is the agency that we rely on to achieve this: provided
that the structure of a star is not strongly dependent on the velocity field
within it, viscous dissipation should ensure that a non-uniformly rotating
star evolves towards its minimum-energy state (for a given angular momentum)
of uniform rotation. Although tidal friction {\it between} the two components
of system A in SS Lac is probably negligible, tidal friction {\it within}
either Aa or Ab is expected to operate on the timescale $t_{\rm visc}$ of
equation (9), i.e. decades. Thus it seems quite possible that the star is indeed kept
fairly near a state of uniform rotation, despite major changes in the direction of
its rotation axis.

   V907 Sco (B9.5V + B9.5V; 3.78d, $e$=0; Lacy, Helt \& Vaz 1999) is another system 
in which eclipses come and go, even more dramatically than in SS Lac. It eclipsed 
in the intervals 1899 -- 1918 and 1963 -- 1986, and not in 1918 -- 1963, or after 
1986. Lacy \etal (1999) detected the third body, also from the motion of the CG of 
the short-period pair, with $P_{\rm out} = 99.3\thin$d, $e_{\rm out} \ts 0$.
Unfortunately an analysable light-curve for this system, during its eclipsing phase,
does not exist, for reasons mentioned by Lacy \etal (1999), and thus we have less 
rather than more data with which to test our model.
\section{Discussion}
Our formulation of the combined effect of five different perturbations
on a Keplerian orbit, while very simple, appears to give physically believable 
results in a number of cases. It is however not easy to find observational data
on stellar orbits which will seriously test the model. A major uncertainty is 
the viscous timescale of a star. One can question whether anything so simple
as a unique viscous timescale can be adequate. However the timescale estimated from 
first principles in the Appendix seems surprisingly reasonable for the radio
pulsar system \smc.

    The model gives a determination of the orientation of the outer orbit in
the triple system SS Lac, and though not over-constrained by data currently
available may be challenged by data that should be available in a few decades.
However in this system the quadrupole distortions of the stars are sufficiently
insignificant that only the third-body terms are being tested here.

    A potentially significant statistical effect is predicted on the basis of
the combination of tidal friction with Kozai cycles. We have argued that if the
outer orbit in a triple is moderately highly inclined to the inner, then the inner 
orbit is likely to be shrunk to a limiting value of only 2 or 3 days, supposing 
that it `started' at a longer period. This limiting period will depend on the outer 
orbital period, and also on the rotation rate of the stars, etc. Roughly, it is 
dictated by the fact that for substantial Kozai cycles we need $C \tgs Z$, from 
equations (12) -- (15). This would give a longer limiting period for systems with
a longer $P_{\rm out}$ than $\beta$ Per. On some models of triple star formation, 
inclinations larger than $60\deg$ are as likely as not, and so we might expect a 
significant deficit of orbits above some value in triples, relative to those 
in binaries. Tokovinin (p.c. 1998) has noted that in his multiple-star catalog
(Tokovinin 1997) the distribution of periods among spectroscopic binaries that
are in triples tends to drop off above $5\thin$d, whereas among those that are
not in triples it continues to rise.

We have not yet been able to incorporate in the model a satisfactory approximation
for what we believe is a very important further perturbation to binary orbits:
the effect of ML and AML such as is likely to be experienced by cool stars with
active dynamos in their outer convection zones. If a star is subject to spherically
symmetric ML, the mass lost carries of orbital angular momentum as well as spin
angular momentum: the amount of the latter may be enhanced if there is magnetic
linkage between the star and the wind out to some substantial Alfv{\'e}n radius.
However ML is unlikely to be very spherically symmetric. Also some of the wind may 
well be accreted by the companion star; and furthermore during the accretion process
there is often found to be further ML, and presumably AML, in the form of bipolar
jets from the inner portion of the accretion disc. A variety of possible models
for the ML/AML process can be thought of, but the physical process that they attempt to
model may be too dependent on the details of how the gas actually travels from
one star either to the other, or to infinity, to admit even at first order a
simple yet credible formulation. We hope to attempt this in the future.

An example where this may well be important is the young and active binary
BY Dra (K1Ve + K1Ve; $6\thin$d, $e=0.5$; Vogt \& Fekel 1979). One of the two components shows
rotational modulation with a period of $4\thin$d. This is too slow for
pseudo-synchronism, which at such high eccentricity implies a rotation period of $2\thin$d.
A possible answer is that the component is in a state of transient equilibrium between
magnetic braking, which would tend to slow it down, and pseudo-synchronisation, which
would tend to speed it up. It is by no means improbable that these two timescales
are comparable in this system.

    Both the concepts of tidal distortion and of tidal friction will we hope be
testable with some three-dimensional numerical modeling of stellar interiors that is 
currently being developed: the DJEHUTY project. This project aims at applying existing 
and well-tested 3-D hydrodynamic and thermodynamic (but non-self-gravitating) 
grid-based algorithms to the self-gravitating situation of stellar interiors,
using massively parallel hardware. Although the resolution currently aimed for,
of $\ts10^8$ cells, would not be enough to resolve the surface layers of tidally-distorted
stars, it may well be adequate to resolve the interiors and so determine whether dissipation 
in convective cores may be an effective agent of tidal friction, as we suggest here.

\acknowledgements

This work was undertaken as part of the DJEHUTY project at LLNL.  Work performed at LLNL 
is supported by the DOE under contract W7405-ENG-48.

\appendix
The analysis that gives (a) the extra forces due to the five perturbative processes
listed at the beginning of \S1, and (b) their effect on the orthogonal triad
$\e,\thin\q,\thin\h$, is largely taken from EKH98, except for the third-body
perturbation which was described in KEM98, and GR which is well-known. However EKH98 contains a mistake
of a factor of 2 in the part of the gravitational potential that is due to the
distortion of the stars by their mutual gravitational interaction. We are
grateful to Dr. R. Mardling for pointing this out. We list here the equations
of EKH98 that have to be changed: (36), (38), (75), (76) and (97).
In each of these, the last term on the right, i.e. the term which does {\it not}
involve $\Omega$, should be divided by two. This has no effect on the overall analysis
in that paper. Equation (12) of the present paper, based on equation (97) of
EKH98, contains the correction.

      Equations (10) and (11) of the present paper are obtained from equations
(88) -- (96) of EKH98. In EKH98 expressions were given for the rate of change of
the Euler angles giving the orientation of the $\e,\q,\h$ frame relative to an
inertial frame. In fact the $X,Y,Z$ terms given here are what emerge more directly
from the analysis; though not given explicitly in EKH98 they can be recovered from 
the formulae for rates of change of Euler angles given there.

The the timescale $\tTF$ for tidal friction that we use in this paper has 
been redefined to be twice the value that was used in EKH98.

    A novel result of EKH98 was a determination, exact to the order we work to 
here, of the velocity field in a rotating star which, in the frame that rotates
with the star, suffers a time-dependent tidal perturbation due to the presence of the
other star. Time dependence can arise because the orbit is elliptical, and/or not 
in corotation with the star. We summarise the result here.

In the \et\ model, we approximate that the
density (as well as pressure) is constant on equipotential surfaces of the
instantaneous gravitational field of the companion. This means that 
$$\rho\equ \rho(\rbar)\w,\w \rbar\equ r+r\alpha(r) P_2(\cos\theta)\ ,  \eqno(A1)$$
where $\alpha(r)$ is a dimensionless function that gives the ellipticity of
an equipotential as a function of distance from the centre. The angle $\theta$
is measured from the direction of $\bd(t)$, the separation of the two stellar
centres. Radau's equation
$$ \alpha^{\prime\prime}-{6\alpha\over r^2}+{8\pi r^3\rho(r)\over m(r)}\left( 
{\alpha^{\prime}\over r}+{\alpha\over r^2}\right)\equ 0\w, \eqno(A2)$$
gives $\alpha(r)$, apart from a multiplicative factor which gives 
$\alpha\equ\alpha_1$ at the surface $r=R_1$, where
$$ \alpha_1\equ -\thin{M_2R_1^3\over M_1d^3}{1\over 1-Q}\w. \eqno(A3)$$
$Q$ is related to the classical apsidal motion constant $k_2$: 
$$k_2\eq {1\over 2}\thin{Q\over 1-Q}\w.\eqno(A4)$$

Let
$$F(r,\theta)\eq r^2P_2(\cos\theta)\equ {3\over 2}(\k.\r)^2-{1\over 2}r^2\w,  
\eqno(A5)$$
where $\k\equiv\bd/d$. Since \bd\  is time-varying, both $\alpha$ and \k\  depend 
on $t$, the former because $\alpha\propto 1/d^3$ -- equation (A3). Then with $\rho$ 
as a function of $\rbar$ only, and $\rbar$ viewed as a function of $\r,t$, we obtain
$${\partial\rho\over\partial t}\equ {d\rho\over d\rbar}\thin{\partial\rbar\over
\partial t}\equ {d\rho\over d\rbar}\left({\partial\alpha\over\partial t}{F\over r}+
{3\alpha G\over r}\right)\equ {3\alpha\over r}\thin{d\rho\over d\rbar}
\left(-{1\over d}{\partial d\over\partial t}\thin F+ G\right)\ ,  \eqno(A6)$$
where
$$G(r,\theta)\eq{1\over 3}{\partial F\over\partial t}\equ  \k.\thin\r\ {\partial\k
\over\partial t}.\thin\r\ \w.  \eqno(A7)$$
$F$ and $G$ are obviously orthogonal harmonic functions of degree 2. 

Now consider the velocity field given by
$$\vbf\equ {3\alpha_1\over 2}\beta(r)\left({1\over d}{\partial d\over\partial t}\thin
\nabla F-\nabla G\right)\ .\eqno(A8)$$
Then
$$ \nabla.\rho\vbf\equ{3\alpha_1\over r}\thin{d\thin \rho\beta\over dr}\left(
{1\over d}{\partial d\over\partial t}\thin F- G\right)\ .\eqno(A9)$$
and we can see that equation (A1) is satisfied to first order, provided that
$${d\thin\rho\beta\over dr}\equ {\alpha\over\alpha_1}\thin{d\rho\over dr}\w,\w\ie\w 
\beta\equ {1\over\rho\alpha_1}\int_{R_1}^r\thin\alpha{d\rho\over dr}\thin dr\w.  
\eqno(A10)$$
The lower limit in the integral comes from the boundary condition that the
outer surface ($\rho=0$) is a surface which moves with the fluid, so that the 
velocity must be finite there despite the vanishing density. The function 
$\beta(r)$ is determined unambiguously by the structure of the star, via 
equation (A3) determining $\alpha(r)$, and is well-behaved ($\beta\to 1$) for 
polytropic $(0\tl n\tl 5$) surfaces as $\rho\to 0$, despite the apparent 
singularity there.

\par Using suffices,
$$v_i\equ {3\alpha_1\over 2}\thin\beta(r)s_{ij}x_j\w,\w {\rm where}\w s_{ij}\ \equiv\ 
{1\over d}{\partial d\over\partial t} \thin(3k_ik_j-\delta_{ij})-k_i{\partial k_j
\over\partial t}-{\partial k_i\over\partial t}k_j\w. \eqno(A11)$$
The rate-of-strain tensor is now seen to be
$$t_{ij}\ \equiv\ {\partial v_i\over\partial x_j}+{\partial v_j\over\partial 
x_i}\equ {3\alpha_1\over 2}\thin \left(2\beta s_{ij}+{\beta^{\prime}\over r}\{s_{ik}
x_kx_j+s_{jk}x_kx_i\}\right)\ .  \eqno(A12)$$
We square this and average it over an equipotential (which at this level of
approximation can be taken to be spherical), to obtain
$${1\over 4\pi}\int t_{ij}^2 d\Omega\equ 9\alpha_1^2\ s_{ij}^2\ \left(\beta^2+
{2\over 3}r\beta\beta^{\prime}+{7\over 30}r^2\beta^{\prime 2}\right)\ .  
\eqno(A13)$$

Now,
$$s_{ij}^2\equ 6\left({1\over d}{\partial d\over\partial t}\right)^2 +2 \left( 
{\partial \k\over\partial t}\right)^2\equ {2\over d^2}\left[2\left({\partial 
d\over \partial t}\right)^2+\left({\partial\bd\over\partial t}\right)^2\right]
\ ,  \eqno(A14)$$
and so the rate of dissipation of mechanical energy is
$$\EEdot\equ -{1\over 2}\int\thin\rho wl\thin t_{ij}^2\thin dV\hskip 4.2truein$$
$$\hskip 0.5truein\equ -{9\alpha_1^2\over d^2}
\thin \left[2\left({\partial d\over \partial t}\right)^2+\left({\partial\bd\over
\partial t}\right)^2\right]\int_0^{M_1}wl\thin\left(\beta^2+{2\over 3}r\beta
\beta^{\prime}+{7\over 30}r^2\beta^{\prime 2}\right)\thin dm\ .  \eqno(A15)$$
The parameters $w(r),l(r)$ are the mean velocity and mean free path of turbulent  
eddies. The $\beta$-dependent weight factor in parentheses in equation (A15) is 
what we call 
$\gamma(r)$, and its average over the turbulent convective region of the star is 
the $\overline{\gamma}$ of equation (9). The factor in square brackets in equation
(A15) leads to a functional form of the tidal friction force, as it depends on 
(variable) separation $\bd$, which is the same as the result usually obtained
by arguing that the tidal bulge lags the line-of-centres by some small fixed
amount. Averaged over a Keplerian orbit, it gives $V$ and $W$ -- equations (5)
and (6), and those parts of the terms $X,Y,Z$ in equations (10) -- (12) that arise 
from tidal friction. The details are given in EKH98.

Although a common approximation for $\alpha(r)$ is $\alpha\propto r^3$, and it
is commonly argued from this that, in effect, $\beta\propto r^4$ and $\gamma
\propto r^8$, none of these approximations is at all reliable. EKH98 integrated
two polytropic models, and two MS stellar models. In the $n=3$ polytrope, it
was found (EKH98, Fig 1) that $\alpha$ and $\beta$ decrease by a factor
of about 10, from the surface right to the centre. At the surface $\beta$ is
unity, and $\gamma$ therefore somewhat larger $(\ts 4$). In the
central one-third by radius, roughly the region of convection in an upper
MS star, $0.01\tls\gamma\tls0.03$. We therefore feel that
an estimate of a mean $\gamma\ts 0.01$ is reasonable, for MS stars which are
typically slightly more centrally condensed than an $n=3$ polytrope.

The approximation $\alpha\propto r^3$ is appropriate to the outer layers of a 
star, where the density is low compared with the mean density: in this case
equation (A2) gives $m^{\prime}=0$, i.e. $\rho=0$. It is easy to integrate
equation (A10) by parts, in this case, and the central value of $\beta$ turns
out to be just the ratio of mean density to central density. On the other hand,
$\alpha=$const. gives $\beta=1$ throughout. The truth lies somewhere in between, with
$\alpha\ts$const. near the centre and $\alpha\ts r^3$ in the outer layers.

\clearpage

\begin{table}
\caption{Proto-Algol binaries with different initial periods.}
\begin{center}
\begin{tabular}{cccc}
\tableline
\tableline
$P_{\circ}$ (d)& $P_{\rm end}$ (d)& $T_{\rm circ}$ (yrs)& $i_{\circ}$ (deg)\\
\tableline 
3&2.8&3$\times$10$^7$&99.8 \\
5&1.9&1$\times$10$^7$&97.3 \\
10&1.7&8$\times$10$^5$&95.0 \\
15&0.8&9$\times$10$^4$&94.0 \\
\tableline
\end{tabular}
\end{center}
\end{table}

\begin{table}
\caption{System parameters for SS Lac.}
\begin{center}
\begin{tabular}{cllllllll}
\tableline\tableline
param &obs&guess &param &obs&comp &param &obs&comp\\
\tableline 
$ P\AB$         &679$\thin$d    &          &$\dfII$     &-0.072       &            &$ R\Aa/a\A$&.0741        &            \\
$ e\AB$         &.159           &          &$\bJ$       &             &101.2$\deg$ &$ R\Ab/a\A$&.0715        &            \\
$ P\A$          &14.416$\thin$d &          &$ i\AB$     &             &75.7$\deg$  &$ a\A$     &             &44.7$\Rsun$ \\
$\aJ\eq i\A$    &87.6$\deg$     &          &$ f\Aa\p$ &2.56$\Msun$  &            &$ R\Aa$    &             &3.36$\Rsun$ \\
$\aJ\p\eq i\A\p$&               &73$\deg$  &$ f\Ab\p$ &2.49$\Msun$  &            &$ R\Ab$    &             &3.20$\Rsun$ \\
$ e\A$          &               &.115      &$ f\B\p$  &0.22$\Msun$  &            &$ e\A\p$   &.136         &.138        \\
$\aH$           &               &29$\deg$  &$ M\Aa$     &             &2.93$\Msun$ &$\bJ\p$    &91.7$\deg$   &91.6$\deg$  \\
$\bH$           &               &37$\deg$  &$ M\Ab$     &             &2.85$\Msun$ &$\Te$      &38$\thin$yr  &37.7$\thin$yr \\    
                &               &          &$ M\B$      &             &.798$\Msun$ &$\aJ\p$    &             &72.9$\deg$  \\    
\tableline
\tablecomments{
$f\Aa\p\equ M\Aa\sin^3 i\A\p$; $ f\Ab\p\equ M\Ab\sin^3i\A\p$; $ f\B\p\equ M\B\sin i\AB\p/(M\Aa+M\Ab+M\B)^{2/3}$.\\
Primed quantities refer to epoch 1998; all others, apart from $\Te$,    
refer to epoch 1912, or to constants.}
\end{tabular}
\end{center}
\end{table}

\clearpage
\centerline{\psfig{figure=f1.eps,height=3.2in,bbllx=50pt,bblly=520pt,bburx=560pt,bbury=780pt,clip=}}
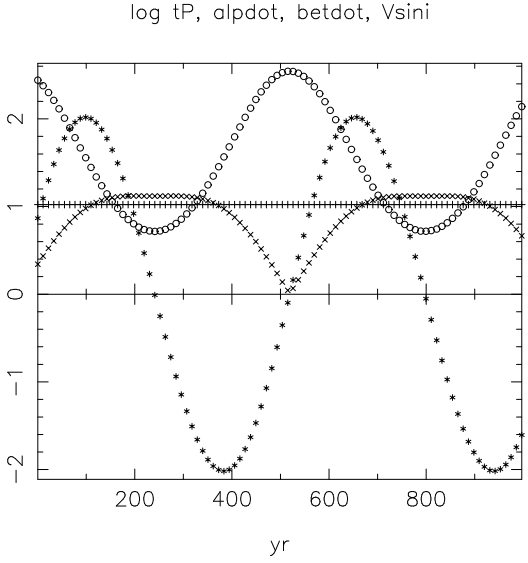
{ Fig 1 -- The evolution of $\log(P/\vert\Pdot\vert)$ (plusses), $\alpdot$ (asterisks), 
$\betdot$ (circles) and $\vr \sin i$ (crosses) 
with time in a binary like the SMC radio pulsar \smc. Each quantity is divided by the
observational value listed in the text. At about $195\thin$yr, all four quantities
are within about $20\%$ of their observational values. At this point, $\bOb=111\deg$, 
having started with an arbitrary value of $0\deg$. Such quantities as $P, e, \Omega$
and $\aOb$ have not changed significantly in this short time.}

\clearpage
\centerline{\psfig{figure=f2.eps,height=6.0in,bbllx=30pt,bblly=260pt,bburx=560pt,bbury=790pt,clip=}}
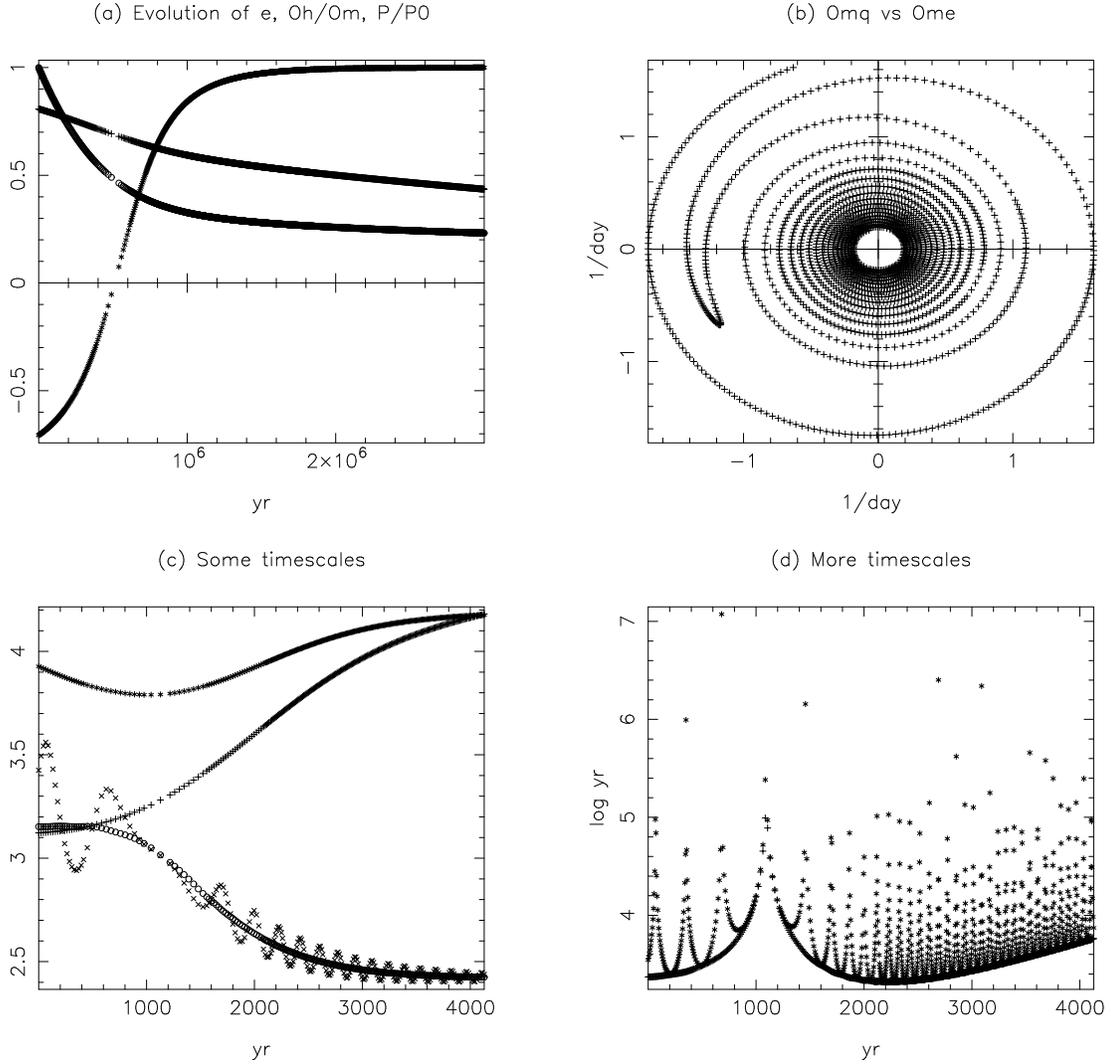
{ Fig 2 -- The evolution of orbit and spin in a binary like the SMC radio
pulsar \smc. (a) Eccentricity (plusses), cosine of the inclination 
of the stellar spin vector $\vO_2$ to the instantaneous orbital plane vector
$\h$ (asterisks), and period relative to initial period (circles). The B star
was started with spin inclined at $\aOb=135\deg$ to the orbit ($\cos\aOb=-0.71$),
reached inclination $90\deg$ after $\ts 5\times 10^5\thin$yr, and was
almost completely parallelised by $\ts 1.7\times 10^6\thin$yr.
(b) The two components of B-star spin in the orbital plane, plotted against
each other. The spin axis started at the top, left-of-centre, turned 
through $\ts 1.25$ rotations anticlockwise, then (at the time when the spin was 
exactly perpendicular to the orbit) reversed its motion to clockwise while 
spiralling in towards the centre. The evolution was speeded up by artificially
decreasing the viscous timescale by a factor of $\ts 500$, to prevent the spiral
being very tightly wound. The `real' timescale would have required about 600
turns before reversal.
(c) Timescales $e/\vert\edot\vert$ (asterisks) and $P/\vert\Pdot\vert$ (plusses); 
also the timescales $1/Z$ (circles) and $1/\vert\betdot\vert$ (crosses) of
apsidal motion (all logged).  The first two timescales are artificially shortened 
by $\ts 500$, as in (b). The last two timescales tend towards equality as the spin
becomes parallelised, but precession due to non-parallel spin causes one to
oscillate about the other. 
(d) The precessional timescales $1/\sqrt{X^2+Y^2}$ (plusses), and the timescale 
$1/\vert\alpdot\vert$ of rate of change of inclination to the line of sight (both
logged). The many cusps in the latter are due to the fact that the inclination was 
oscillating between two values.} 
\clearpage
\centerline{\psfig{figure=f3.eps,height=3.2in,bbllx=50pt,bblly=520pt,bburx=560pt,bbury=780pt,clip=}}
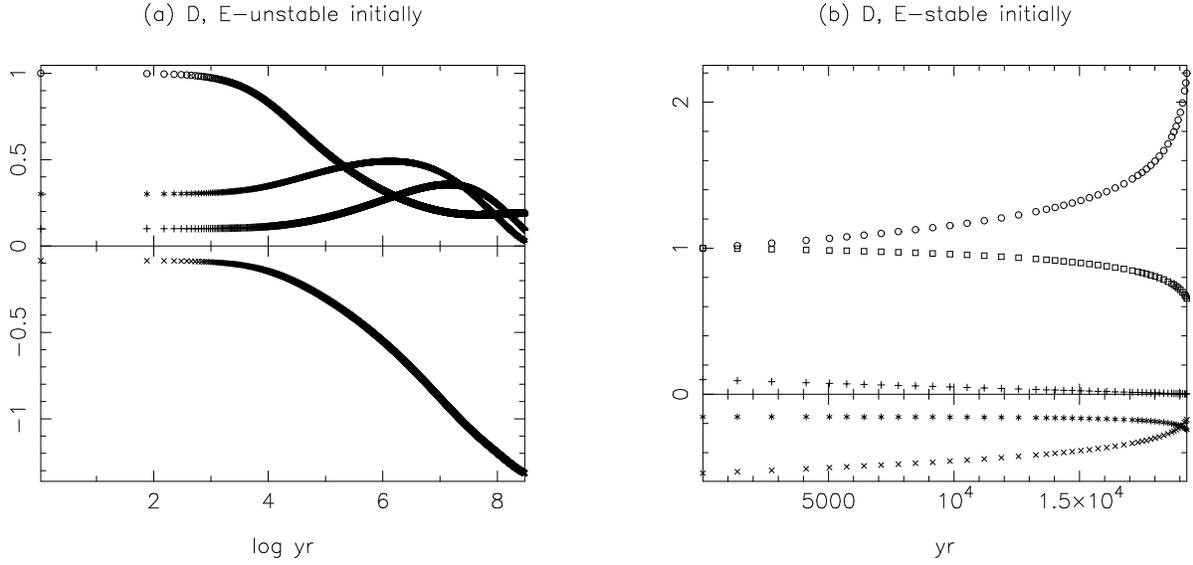
{ Fig 3 --The Darwin (D) and eccentricity (E) instabilities. Eccentricity (plusses),
orbital frequency $\omega$ relative to its initial value (circles), the degree of
asynchronism, $\log(\Omega/\omega)$ (asterisks), and the ratio of spin to orbital
angular momentum, $\log(I\Omega/\mu h)$ (crosses). $*1$ is a neutron star, and
$*2$ a massive, partly-evolved MS star. The orbit has $P=6\thin$d, $e=0.1$ to
start with. (a) Initially $\Omega/\omega=2$. (b) Initially $\Omega/\omega=0.7$. 
In (a) the system starts both D-unstable and E-unstable. Eccentricity and asynchronism
grow, but the periastron separation remains large enough to avoid collision. Once the
orbit has widened it becomes stable to both processes, and settles down. However
nuclear evolution (neglected) would cause problems before $10^7\thin$yr. In (b) the
orbit is E-stable and slightly D-stable to start with. But as the orbit and star gradually
spin up the orbit's angular momentum goes down, while the star's goes up, leading to
D-instability in about 5000$\thin$yr. After that asynchronism increases, and the stars
collide in about $19000\thin$yr.}
\clearpage
\centerline{\psfig{figure=f4.eps,height=2.8in,bbllx=40pt,bblly=520pt,bburx=580pt,bbury=770pt,clip=}}
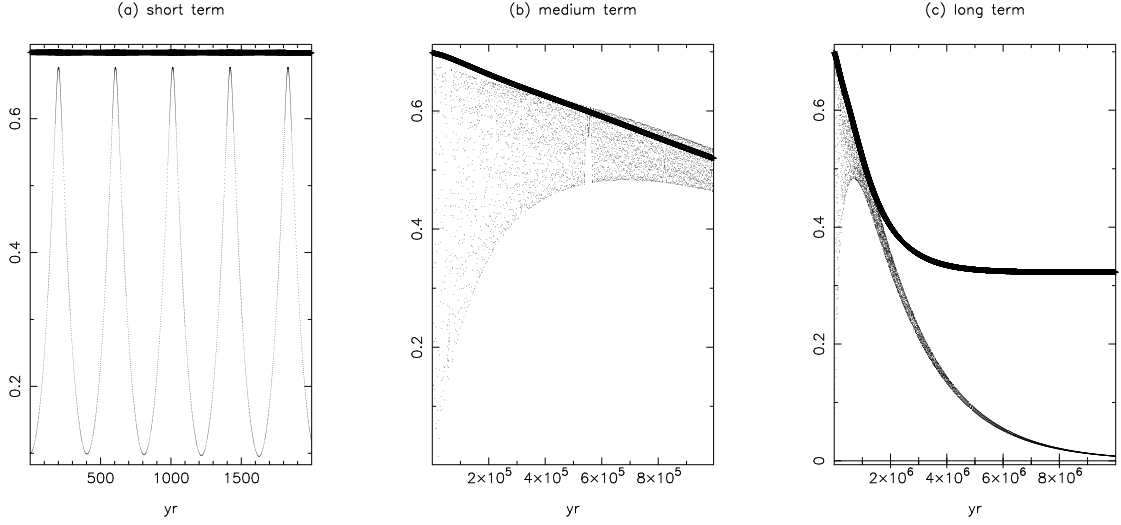
{ Fig 4 -- Evolution of eccentricity (dots) and $\log P$ (thick line) in the 
inner binary of a `proto-Algol' triple system. The initial orbital parameters
are $((2.5 + 2\Msun; 5\thin$d, $e=0.1$) + $1.7\Msun$; $679\thin$d, $e=0.23$; 
$\aH=97.5\deg$). (a) The first $2000\thin$yr, showing somewhat truncated Kozai
cycles; (b) the first $10^6\thin$yr, showing the orbit settling towards a nearly 
constant but slowly decreasing eccentricity; (c) the first $10^7\thin$yr. By 
$10^7\thin$yr, $e\ts 0$, $P\ts 2.1\thin$d, and $\aH=100\deg$. Some apparent 
structure in the eccentricity variation in (b) is due to beating between the 
data-plotting frequency and Kozai-cycle frequency, which can be commensurable.}

\clearpage
\centerline{\psfig{figure=f5.eps,height=3.0in,bbllx=30pt,bblly=525pt,bburx=580pt,bbury=770pt,clip=}}
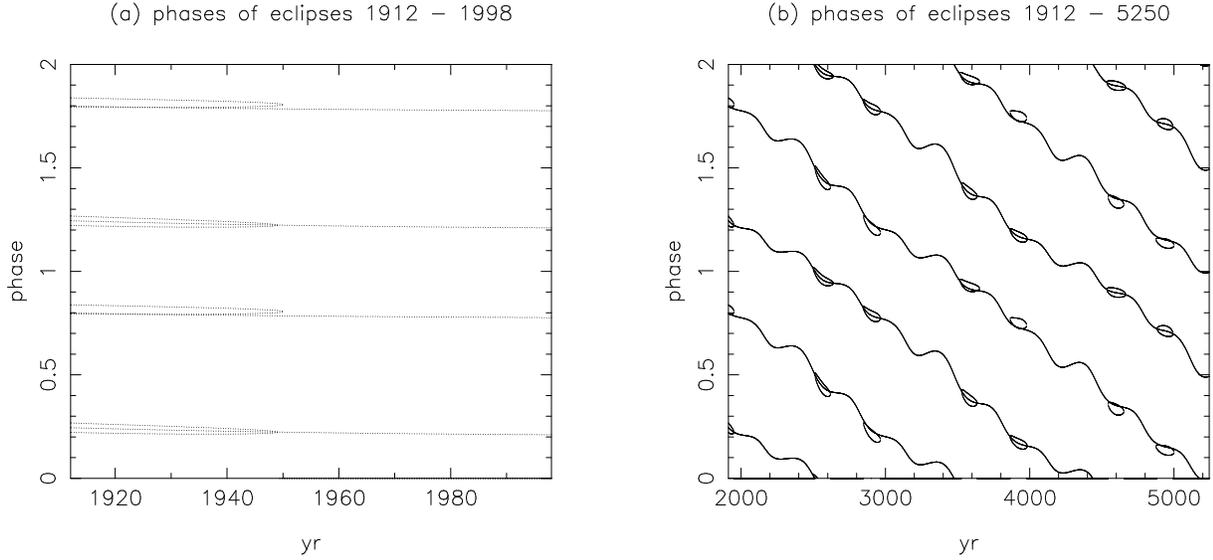
{ Fig 5 -- (a) The phases of eclipses of SS Lac as computed here for 1912 - 1998. 
Two whole cycles are shown on the vertical axis. Phase zero is periastron. Eclipses 
occur in the narrow cigar-shaped areas centred at phases 0.24 (eclipse II, $*1$ in
front) and 0.81 (eclipse I, $*1$ behind), and ending at about 1950. Also shown are 
two phases labelled (i) and (ii). The first is where $*1$ crosses the plane containing 
the line of sight and the orbital axis, $*1$ in front -- equation (32) -- and the 
second where the radial velocity of $*1$ relative to the CG of the inner binary is 
zero and decreasing (\ie\  $*1$ behind) -- 
equation (34). The slight effect on phase of the orbital motion of the inner binary 
within the outer binary has been ignored. (b) Same as (a) but for the interval 1912
-- 5250. Regions of eclipses are now leaf-shaped. The sloping lines can be identified 
by comparing the left-hand edge with the whole of (a). The slopes indicate that
periastron is, on the whole, advancing, but occasionally retreats because of
precession.}

\clearpage
\centerline{\psfig{figure=f6.eps,height=2.8in,bbllx=30pt,bblly=270pt,bburx=580pt,bbury=525pt,clip=}}
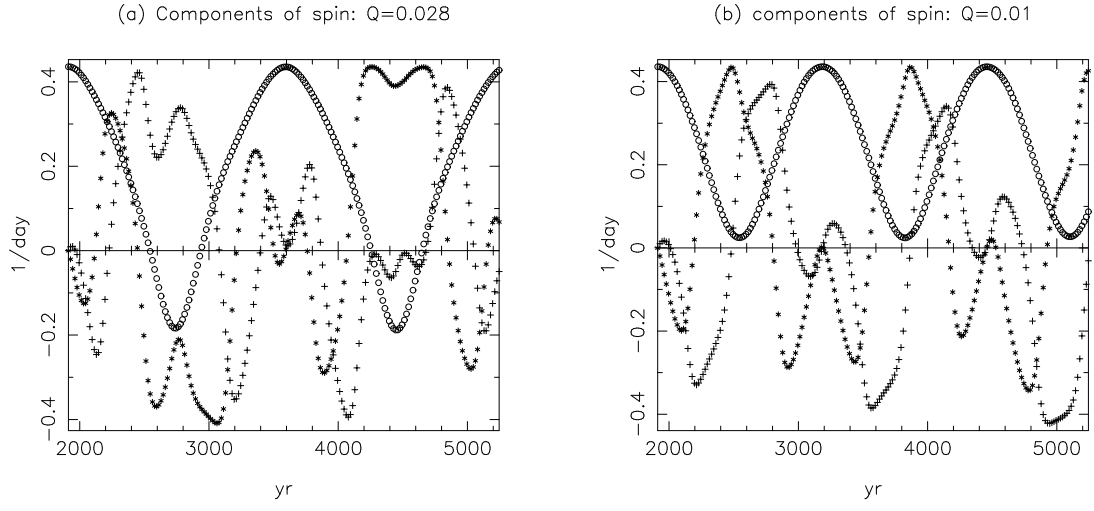
{ Fig 6 -- The three components of the angular velocity of $*1$ in SS Lac as 
functions of time: $\Oah$ (circles), $\Oae$ (plusses) and $\Oaq$ (asterisks). 
(a) $Q_1=Q_1=0.028$; (b) $Q_1=Q_2=0.01$. In both cases, the system was started with
parallel corotation.}
\clearpage


\begin{thebibliography}{}
\bibitem[]{} Alexander, M. E. (1973) Ap. Sp. Sc., 23, 459 
\bibitem[]{} Bell, J. F., Bessell,  M. S., Stappers, B. W., Bailes, M. \& 
            Kaspi, V. M. (1995) ApJ, 447, L117
\bibitem[]{} Claret, A \& Gim{\'e}nez, A. (1993) A\&A,277, 487 
\bibitem[]{} Eggleton, P. P. (2000a) New Astronomy Reviews, 44, 11
\bibitem[]{} Eggleton, P. P. (2000b) Proceedings of Bormio workshop `{\it
Evolutionary processes in Binary and Multiple Systems}', eds. Podsiadlowski, Ph. \&
Rappaport, S. A.\  ASP Cons. Ser. to be published
\bibitem[]{} Eggleton, P. P., Kiseleva, L. G. \& Hut, P. (1998; EKH98) ApJ, 499, 853
\bibitem[]{} Guinan, E. F., Marshall, J. J. \& Maloney, D. (1994) IBVS 4101 
\bibitem[]{} Hut, P. (1981) A\&A, 99, 126
\bibitem[]{} Iben, I., Jr \& Tutukov, A. V. (1998; IT) ApJ, 501, 263
\bibitem[]{} Kaspi, V. M., Johnston, S., Bell, J. F., Manchester, R. N., Bailes, M., Bessell, M., 
            Lyne, A. G. \& D'Amico, N. (1994) ApJ, 423, L43 
\bibitem[]{} Kaspi, V. M., Bailes, M., Manchester, R. N., Stappers, B. W. \& Bell, J. F. (1996) 
            Nature, 381, 584 
\bibitem[]{} Kiseleva, L. G., Eggleton, P. P. \& Mikkola, S. (1998; KEM98) MN, 300, 292
\bibitem[]{} Kozai, Y. 1962 AJ, 67, 591
\bibitem[]{} Lacy, C. H. S.  (1997) AJ, 115, 801 
\bibitem[]{} Lacy, C. H. S., Helt, B. E. \& Vaz, L. P. R. (1999) AJ, 117, 541
\bibitem[]{} Lestrade, J.-F., Phillips, R. B., Hodges, M. W. \& Preston, R. A. (1993)
            ApJ, 410, 808 
\bibitem[]{} Mazeh, T. \& Shaham, J. (1979) A\&A, 77, 145 
\bibitem[]{} Refsdal, S., Roth, M. L. \& Weigert, A., 1974, A\&A, 36, 113 
\bibitem[]{} Schmutz, W., Schweickhart, J., Stahl, O.,Wolf, B., Dumm, T., G{\"a}ng, Th., Jankovics,
 I., Kaufer, A., Lehmann, H., Mandel, H., Peitz, J. \& Revinius, Th. (1997) A\&A, 328, 219
\bibitem[]{} S{\"o}derhjelm, S. (1975) A\&A, 42, 229
\bibitem[]{} Tokovinin, A. A. (1997) A\&AS, 124, 75) 
\bibitem[]{} Torres, G. \& Stefanik, R. P. (2000; TS00) AJ, 119, 1914 
\bibitem[]{} van den Heuvel, E. P. J. \& van Paradijs, J. (1997; HP) ApJ, 483, 399
\bibitem[]{} Vogt, S. \& Fekel, F. C. (1979) ApJ, 234, 958
\bibitem[]{} Witte, M. \& Savonije, G. T. (1999) A\&A, 350, 129 
\bibitem[]{} Zahn, J.-P. (1977) A\&A, 57, 383 
\bibitem[]{} Zahn, J.-P. (1978) A\&A, 67, 162 
\end{thebibliography}
\end{document}